\documentclass[%
superscriptaddress,
preprint,
preprintnumbers,
 amsmath,amssymb,
 aps,
notitlepage,
showpacs,
]{revtex4-2}
\usepackage{graphicx}% Include figure files
\usepackage{float}
\usepackage{hyperref}
\usepackage{booktabs}
\usepackage{multirow}
\usepackage{siunitx}
\usepackage{parskip}
\usepackage{dcolumn}% Align table columns on decimal point
\usepackage{bm}% bold math
\newcommand{\Tr}{{\rm Tr}}
\newcommand{\tr}{{\rm tr}}
\newcommand{\re}{{\rm Re}}
\newcommand{\Dodwf}{\mathcal{D}}
\newcommand{\Id}{\mbox{1\hspace{-1.2mm}I}}
\newcommand{\DWF}{{\rm DWF}}
\newcommand{\ov}{{\rm ov}}

\newcommand{\clover}{{\rm clover}}
\newcommand{\fm}{{\rm fm}}
\newcommand{\round}{{\rm round}}

\newcommand{\bea}{\begin{eqnarray}}
\newcommand{\eea}{\end{eqnarray}}
\newcommand{\BAN}{\begin{eqnarray*}}
\newcommand{\EAN}{\end{eqnarray*}}
\newcommand{\nn}{\nonumber\\}

\begin{document}

\newcommand{\ASIOP}
{Institute of Physics, Academia Sinica, Taipei, Taiwan~11529, R.O.C.}

\newcommand{\NTU}
{Physics Department, National Taiwan University, Taipei, Taiwan~10617, R.O.C.}

\newcommand{\NTNU}
{Physics Department, National Taiwan Normal University, Taipei, Taiwan~11677, R.O.C.}

\newcommand{\RCAS}
{Research Center for Applied Sciences, Academia Sinica, Taipei, Taiwan 11529, R.O.C.}

\preprint{NTUTH-22-505A}

\title{Topological susceptibility in finite temperature QCD 
       with physical $(u/d, s, c)$ domain-wall quarks}

\author{Yu-Chih~Chen}
\affiliation{\NTU}

\author{Ting-Wai~Chiu}
\email{twchiu@phys.ntu.edu.tw}
\affiliation{\NTU}
\affiliation{\ASIOP}
\affiliation{\NTNU}

\author{Tung-Han~Hsieh}
\affiliation{\RCAS}

\collaboration{for the TWQCD Collaboration} %\noaffiliation

\begin{abstract}

We perform hybrid Monte-Carlo (HMC) simulation 
of lattice QCD with $N_f=2+1+1$ domain-wall quarks at the physical point,
on the $64^3 \times (64,20,16,12,10,8,6)$ lattices, each with three lattice spacings.  
%($a \sim \{0.064, 0.068, 0.075\}$~fm). 
The lattice spacings and the bare quark masses are determined on the $64^4$ lattices.
The resulting gauge ensembles provide a basis for studying finite temperature QCD
with $N_f=2+1+1 $ domain-wall quarks at the physical point. 
In this paper, we determine the topological susceptibility of the QCD vacuum 
for $T > T_c \sim 150 $~MeV. The topological charge of each gauge configuration
is measured by the clover charge in the Wilson flow at the same flow time in physical units, 
%$ t = 0.8192~{\rm fm}^2 $, 
and the topological susceptibility $ \chi_t(a,T) $ 
is determined for each ensemble with lattice spacing $a$ and temperature $T$.  
Using the topological susceptibility $\chi_t(a,T) $ of 15 gauge ensembles
with three lattice spacings and different temperatures in the range $T \sim 155-516 $~MeV,
we extract the topological susceptibility $\chi_t(T)$ in the continuum limit.
To compare our results with others, we survey the continuum extrapolated 
$\chi_t(T)$ in lattice QCD with $N_f=2+1(+1)$ dynamical quarks at/near the physical point, 
and discuss their discrepancies.
Moreover, a detailed discussion on the reweighting method for domain-wall fermion is presented.

\end{abstract}

%% \tableofcontents

\maketitle

\section{Introduction}
\label{sec:introduction}

The topological susceptibility $ \chi_t $ is the most crucial quantity to measure
the quantum fluctuations of the QCD vaccum, and it is defined as
\begin{equation}
\chi_t = \lim_{V \to \infty} \frac{\langle Q_t^2 \rangle}{V},
\end{equation}
where $ Q_t $ is the integer-valued topological charge of the gauge field
in the 4-dimensional volume $V$,
\begin{equation}
\label{eq:Qt}
Q_t=\frac{g^2 \epsilon_{\mu\nu\lambda\sigma}}{32 \pi^2} \int d^4 x \
\tr[ F_{\mu\nu}(x) F_{\lambda\sigma}(x)],
\end{equation}
and $ F_{\mu\nu} = T^a F_{\mu\nu}^a$ is the matrix-valued field tensor, with the normalization
$ \tr (T^a T^b) = \delta_{ab}/2 $.

At low temperature $ T < T_c \simeq 150 $~MeV, 
$ \chi_t $ is related to the chiral condensate $ \Sigma $,  
\begin{equation}
\Sigma = -\lim_{m_q \to 0} \lim_{V \to \infty} 
                           \frac{1}{\Omega} \int d^4 x \ \langle \bar{q}(x) q(x) \rangle,  
\end{equation}
the order parameter of the spontaneously chiral symmetry breaking, and its nonzero value 
gives the majority of visible (non-dark) mass in the present universe.  

For QCD with $u$ and $d$ light quarks, the chiral perturbation theory (ChPT) at the tree level 
gives the relation \cite{Leutwyler:1992yt}
\begin{equation}
\label{eq:LS}
\chi_t = \Sigma \left( \frac{1}{m_u} + \frac{1}{m_d} \right)^{-1},  
\end{equation}
which shows that $ \chi_t $ is proportional to $ \Sigma $.
This implies that the non-trivial topological quantum fluctuations is the origin 
of the spontaneously chiral symmetry breaking. In other words,   
if $\chi_t $ is zero, then $ \Sigma $ is also zero and the chiral symmetry is unbroken, 
and the mass of neutron/proton could be as light as $\sim 10 $~MeV rather than $\sim 940$~MeV.
Moreover, $\chi_t$ breaks the $ U_A(1) $ symmetry
and resolves the longstanding problem why the flavor-singlet $ \eta'$ is much
heavier than other non-singlet (approximate) Goldstone bosons 
%\cite{tHooft:1976rip,tHooft:1976snw,Witten:1979vv,Veneziano:1979ec}.
\cite{tHooft:1976rip,Witten:1979vv,Veneziano:1979ec}.

At finite temperature $ T < T_c $, for small quark masses,  
%($ m_q \ll \Lambda_{QCD} $),  
the ChPT asserts that $ \chi_t(T) $ is proportional to $ \Sigma(T) $,  
and provides a prediction of $ \chi_t(T) $
with the input $ \chi_t(0) $ at zero temperature 
\cite{Gasser:1986vb,Gasser:1987ah,Gerber:1988tt,Hansen:1990yg}.

At temperature $ T > T_c $, the chiral symmetry is restored and $ \Sigma(T) = 0 $.
However, it is unclear whether $ U(1)_A $ is also restored at $ T_1 \sim T_c $.  
Remarkably, if $ U(1)_A $ is broken up to some $T_1 > T_c $ and restored for $ T > T_1 $, 
then there exists an interval $ (T_c, T_1) $ in which  	
the non-trivial quantum fluctuations of the QCD vacuum can only make  
$ \chi_t(T) $ nonzero but not $ \Sigma(T) $. It is interesting to understand 
the physics underlying this mechanism. 

Moreover, another interesting aspect of $\chi_t(T)$ is that it could  
play an important role in generating the majority of mass in the universe, 
as a crucial input to the axion mass and energy density, a promising candidate 
for the dark matter in the universe. 
The axion 
%\cite{Peccei:1977hh,Peccei:1977ur,Weinberg:1977ma,Wilczek:1977pj} 
\cite{Peccei:1977hh,Weinberg:1977ma,Wilczek:1977pj} 
is a pseudo Nambu-Goldstone boson arising from the breaking of a 
hypothetical global chiral $U(1)$ extension of the Standard Model 
at an energy scale $f_A$ much higher than the electroweak scale, the Pecci-Quinn mechanism.  
This not only solves the strong CP problem, but also provides an explanation 
for the dark matter in the universe.  
The axion mass at temperature $T$ is proportional to $ \sqrt{\chi_t(T)} $,           
\begin{equation}
m_A(T) = \frac{\sqrt{\chi_t(T)}}{f_A}, 
\end{equation}
which is one of the key inputs to the equation of motion for the axion field evolving from 
the early universe to the present one, with solutions predicting the relic axion energy density,  
through the misalignment mechanism \cite{Dine:1981rt,Preskill:1982cy,Abbott:1982af}. 

In general, the determination of $ \chi_t(T) $ requires nonperturbative approaches 
from the first principles of QCD.
To this end, lattice QCD provides a viable nonperturbative determination of $ \chi_t(T) $.
Nevertheless, it becomes more and more challenging as the temperature gets higher and higher,
since in principle the non-trivial configurations are more suppressed at higher temperatures,
which in turn must require a much higher statatics in order to give a reliable determination.
So far, direct simulations have only measured $ \chi_t(T) $ up to $ T \sim 550 $~MeV.
Nevertheless, for $ T \gg T_c $, the temperature dependence of $\chi_t(T) $ can be obtained
with the dilute instanton gas approximation (DIGA), 
which gives $ \chi_t(T) \sim T^{-7-N_f/3} $ for $N_f$ flavors of quarks \cite{Gross:1980br}.

Recent lattice studies of $\chi_t(T) $ aiming at the axion cosmology include
various simulations with $N_f=0$, $2+1$, and $2+1+1$,
where the lattice fermions in the unquenched simulations
include the staggered fermion, the Wilson fermion, and the Wilson twisted-mass fermion
\cite{Berkowitz:2015aua, Kitano:2015fla, Borsanyi:2015cka, Bonati:2015vqz,
      Petreczky:2016vrs, Borsanyi:2016ksw, Burger:2018fvb}.
For recent reviews, see, e.g., Refs. \cite{Moore:2017ond, Lombardo:2020bvn} and references therein.

In this study, we perform the HMC simulation of lattice QCD with $N_f=2+1+1$
optimal domain-wall quarks at the physical point,
on the $64^3 \times (64, 20, 16, 12, 10, 8, 6)$ lattices,
each with three lattice spacings $a \sim (0.064, 0.068, 0.075) $~fm.
The bare quark masses and lattice spacings are determined on the $64^4$ lattices.
The topological susceptibility of each gauge ensemble is measured by the Wilson flow
at the flow time $ t = 0.8192~{\rm fm}^2 $,
with the clover definition for the topological charge $ Q_t $.
Using the topological susceptibility $\chi_t(a,T)$
of 15 gauge ensembles with 3 different lattice spacings
and different temperatures in the range $T \sim 155-516$~MeV,
we extract the topological susceptibility $\chi_t(T)$ in the continuum limit.
Our preliminary results of $ \chi_t(T) $ have been presented in lattice 2021 \cite{Chiu:2021ndv}.

The outline of this paper is as follows.
In Section \ref{section:HMC}, 
we give a description of our HMC simulation with $N_f=2+1+1$ domain-wall quarks at 
the physical point, including the actions, the algorithms, the gauge ensembles, 
the quark propagators, and the residual masses. 
In Section \ref{section:chit}, 
we describe our measurements of the topological susceptibility for our gauge ensembles, 
and the extrapolation to the continuum limit. 
In Section \ref{section:volume}, 
we investigate the volume dependence of the topological susceptibility, 
by comparing the results between two spatial volumes $ \sim (4~\fm)^3 $ and $ \sim (2~\fm)^3 $, 
for $ T \sim 190-510$~MeV.  
In Section \ref{section:compare_overlap_clover}, 
we compare the topological charge/susceptibility of two different definitions:  
the index of the overlap Dirac operator versus the clover charge in the Wilson flow.
In Section \ref{section:reweighting}, we give a detailed discussion on   
the reweighting method for domain-wall fermion.   
In Section \ref{section:compare_chit}, 
we survey the continuum extrapolated topological susceptibility in 
recent lattice studies with $N_f=2+1(+1)$ dynamical fermions at/near 
the physical point, and discuss their discrepancies.
In Section \ref{section:summary}, we conclude with some remarks.
In Appendix A, we present our results of renormalized chiral condensate 
for $T \simeq 131 - 516 $~MeV.  

\section{Simulation of $N_f=2+1+1$ lattice QCD with domain-wall quarks}
\label{section:HMC}

The first HMC simulation of $N_f=2+1+1$ QCD with domain-wall quarks 
was performed on the $32^3 \times 64 $ lattice with physical $m_s$ and $m_c$, 
but unphysical $m_{u/d}$ with $ M_{\pi}^\pm \sim 280 $~MeV \cite{Chen:2017kxr}.
Later the simulation was extended to physical $m_{u/d}$, $m_s$ and $m_c$ on the $64^4 $ lattice, 
with $ a \simeq 0.064$~fm, $ L > 4$~fm and $M_\pi L > 3 $ \cite{Chiu:2018qcp,Chiu:2020ppa}.
Our present simulations with physical $(u/d, s, c)$
on the $64^3 \times (64, 20, 16, 12, 10, 8, 6) \equiv (N_x^3, N_t) $ lattices are extensions 
of our previous ones, using the same actions and algorithms, 
and the same simulation code with tunings for the computational platform Nvidia DGX-V100. 
Most of our production runs were performed on 10-20 units of Nvidia DGX-V100 at two institutions 
in Taiwan, namely, Academia Sinica Grid Computing (ASGC) and 
National Center for High Performance Computing (NCHC), from 2019 to 2021. 
Besides Nvidia DGX-V100, we also used other Nvidia GPU cards 
(e.g., RTX-2080Ti, GTX-1080Ti, GTX-TITAN-X, GTX-1080) for HMC simulations on 
the $64^3 \times (12,8,6) $ lattices, which only require 8-22 GB device memory. 
In the following, we outline our HMC simulations of lattice QCD with $N_f=2+1+1$ 
optimal domain-wall quarks at the physical point. 

\noindent{\bf A. Lattice actions}

For the gluon action, we use the Wilson plaquette action \cite{Wilson:1974sk}
\begin{equation}
S_g(U) = \beta \sum_{\text{plaq.}} \left\{1-\frac{1}{3} \re \Tr (U_p) \right\},  
\end{equation}
where $ \beta = 6/g_0^2 $, and the boundary conditions of the link variables are 
periodic in all directions of the 4-dimensional lattice.   
Then setting $\beta$ to three different values 
$ \beta = \{ 6.15, 6.18, 6.20 \}$ gives three different lattice spacings 
$a \simeq (0.075, 0.068, 0.064) $~fm respectively.
For each lattice spacing, the bare masses of $(m_{u/d}, m_s, m_c)$ are tuned 
such that the lowest-lying masses of the meson operators 
$ \{ \bar{u} \gamma_5 d, \bar{s} \gamma_i s, \bar{c} \gamma_i c \} $ 
are in good agreement with the physical masses of 
$\{ \pi^{\pm}(140), \phi(1020), J/\psi(3097) \}$ respectively. 
%within error bars $\{5, 8, 10\}$~MeV respectively.
  
For the quark action, we use optimal DWF 
with the 5-dimensional lattice fermion operator \cite{Chiu:2002ir}, 
\begin{equation}
\label{eq:D_odwf}
[\mathcal{D}(m_q)]_{xx';ss'}(m_q) =
  (\omega_s D_w + 1)_{xx'} \delta_{ss'}
 +(\omega_s D_w - 1)_{xx'} L_{ss'},
\end{equation}
where $ \{ \omega_s, s = 1, \cdots, N_s \} $ are given by the exact solution 
such that the effective 4-dimensional lattice Dirac operator possesses 
optimal chiral symmetry for any finite $ N_s $,  
i.e., the sign function $S_{N_s}(H_w)$ [see (\ref{eq:Dmq_Ns})] 
is exactly equal to the Zolotarev optimal rational approximation of $ H_w/\sqrt{H_w^2} $.     
The indices $ x $ and $ x' $ denote the lattice sites on the 4-dimensional lattice,
$ s $ and $ s' $ the indices in the fifth dimension,
and the Dirac and color indices have been suppressed.
Here $D_w$ is the standard Wilson Dirac operator plus a negative parameter $-m_0$ 
which is fixed to $-1.3$ in our simulations,  
\begin{equation}
(D_w)_{xx'} = (4-m_0) -\frac{1}{2} \sum_{\hat\mu=1}^4 \left[
  (1-\gamma_\mu)U_\mu(x)\delta_{x+\hat{\mu},x'}
 +(1+\gamma_\mu)U^\dagger_\mu(x')\delta_{x-\hat{\mu},x'} \right],
\end{equation}
where $U_\mu(x)$ denotes the link variable pointing from $ x $ to $ x + \hat\mu $.
The boundary conditions of $ D_w $ on the 4-dimensional lattice 
are periodic in space and antiperiodic in time.
The operator $ L $ is independent of the gauge field, and it can be written as
\begin{eqnarray}
\label{eq:L}
L = P_+ L_+ + P_- L_-, \quad P_\pm = (1\pm \gamma_5)/2,
\end{eqnarray}
and
\begin{eqnarray}
\label{eq:Lpm}
(L_+)_{ss'} = (L_-)_{s's} = \left\{
    \begin{array}{ll}
      - (m_q/m_{PV}) \delta_{N_s,s'}, & s = 1,  \\
      \delta_{s-1,s'}, & 1 < s \leq N_s
    \end{array}\right.,
\end{eqnarray}
where $ m_q $ is the bare quark mass, 
and $ m_{PV} = 2 m_0 $ is the Pauli-Villars mass of optimal DWF.
Note that the matrices $ L_{\pm} $ satisfy $ L_\pm^T = L_\mp $, and $ R_5 L_\pm R_5 = L_\mp $,
where $ R_5 $ is the reflection operator in the fifth dimension,
with elements $ (R_5)_{ss'} = \delta_{s',N_s+1-s} $. 
Thus $ R_5 L_\pm $ is real and symmetric. 

Note that the Pauli-Villars mass $m_{PV}$ is the upper cutoff for the quark mass $m_q$, since 
in the limit $ m_q = m_{PV} $ the theory is reduced to the quenched approximation. 
Thus any quark mass $ m_q $ is required to satisfy $ m_q \ll m_{PV} $. Otherwise, the 
systematic error due to the mass cutoff is out of control.      
In general, the value of $ m_{PV} $ is $ 2 m_0(1-d m_0) $, 
where $ d $ is a parameter depending on the variant of DWF, e.g., 
$d=0$ and $ m_{PV} = 2m_0 < 4 $ for optimal DWF, 
and $ d = 1/2 $ and $ m_{PV} = m_0(2-m_0) < 1 $ for the Shamir/M\"obius DWF. 
Thus optimal DWF has the maximum value of $ m_{PV} = 2 m_0 $,    
and it is theoretically the best choice for the simulation of 
lattice QCD with heavy $c$ and $b$ quarks, 
see Ref. \cite{Chiu:2020tml} for further discussions.

The pseudofermion action for optimal DWF can be written as
\begin{equation}
\label{eq:S_nf1}
S = \phi^\dagger \frac{\mathcal{D}(m_{PV})}{\mathcal{D}(m_q)} \phi, \hspace{4mm} m_{PV}=2 m_0,   
\end{equation}
where $ \phi $ and $ \phi^\dagger $ are complex scalar fields carrying the same quantum numbers
(color, spin) of the fermion fields. 
Integrating the pseudofermion fields in the fermionic partition function 
gives the fermion determinant of the effective 4-dimensional lattice Dirac operator $D_{N_s}(m_q)$, 
i.e., 
\begin{equation}
\label{eq:Z_nf1}
\int [d\phi^{\dagger}][d\phi] \exp\left\{ -\phi^\dagger \frac{\mathcal{D}(m_{PV})}{\mathcal{D}(m_q)} \phi \right\}
= \det \frac{\mathcal{D}(m_q)}{\mathcal{D}(m_{PV})}
= \det  D_{N_s}(m_q), 
\end{equation}
where  
\begin{eqnarray}
\label{eq:Dmq_Ns}
\begin{aligned}
D_{N_s}(m_q) &= m_q + \frac{1}{2}(m_{PV}-m_q) [1+ \gamma_5 S_{N_s}(H_w) ], 
\hspace{2mm} H_w=\gamma_5 D_w  \\
S_{N_s}(H_w) &= \frac{1-\prod_{s=1}^{N_s} T_s}{1 + \prod_{s=1}^{N_s} T_s}, 
\hspace{4mm} T_s = \frac{1-\omega_s H_w}{1+ \omega_s H_w}.
\end{aligned}
\end{eqnarray}
Note that the counterpart of Eq. (\ref{eq:Dmq_Ns}) for Shamir/M\"obius DWF can be obtained 
by replacing $ H_w $ with $ H=\gamma_5 D_w (2 + D_w)^{-1} $, $ m_{PV} = m_0 ( 2 - m_0 ) $,   
and setting $ \{ \omega_s = 1, s=1, \cdots, N_s \}$. 

In the limit $ N_s \to \infty $, $ S_{N_s}(H_w) \to H_w/\sqrt{H_w^2} $, 
and $ D_{N_s}(m_q) $ goes to 
\begin{equation}
\label{eq:Dmq_overlap}
D_\ov(m_q) = m_q + \frac{1}{2} \left( m_{PV} - m_q \right) \left[1+ \gamma_5 S(H_w) \right], 
\hspace{4mm} S(H_w) \equiv \frac{H_w}{\sqrt{H_w^2}}. 
\end{equation}
In the massless limit $ m_q = 0 $, $D_\ov(0)$ is equal to the 
overlap-Dirac operator \cite{Neuberger:1997fp},  
and it satisfies the Ginsparg-Wilson relation \cite{Ginsparg:1981bj}
\begin{eqnarray}
\label{eq:GW}
D_\ov(0) \gamma_5 + \gamma_5 D_\ov(0) = \frac{2}{m_{PV}} D_\ov(0) \gamma_5 D_\ov(0) \iff
D_\ov^{-1} \gamma_5  +  \gamma_5 D_\ov^{-1} = \frac{2}{m_{PV}} \gamma_5 \Id,
\end{eqnarray}
where the chiral symmetry is broken by a contact term, 
i.e., the exact chiral symmetry at finite lattice spacing.  

For finite $N_s$, the exact chiral symmetry is broken, 
but optimal chiral symmetry can be attained if $ S_{N_s}(H_w) $ is equal to the 
Zolotarev approximation of the sign function $ H_w/\sqrt{H_w^2} $, which 
can be achieved by fixing $ \{ \omega_s \} $ according to the exact solution \cite{Chiu:2002ir}, 
\begin{equation}
\label{eq:omega}
\omega_s = \frac{1}{\lambda_{min}} \sqrt{ 1 - \kappa'^2 \mbox{sn}^2
                  \left( v_s ; \kappa' \right) }, \hspace{4mm} s = 1, \cdots, N_s,
\end{equation}
where $ \mbox{sn}( v_s; \kappa' ) $ is the Jacobian elliptic function
with argument $ v_s $ and modulus $ \kappa' =\sqrt{ 1 - \lambda_{min}^2 / \lambda_{max}^2 } $.   
Then $ S_{N_s}(H_w) $ is exactly equal to the 
Zolotarev optimal rational approximation of $ H_w/\sqrt{H_w^2} $, 
i.e., the approximate sign function $ S_{N_s}(H_w) $  
satisfying the bound $ | 1-S_{N_s}(\lambda) | \le d_Z $ 
for $ \lambda^2 \in [\lambda_{min}^2, \lambda_{max}^2] $,  
where $ d_Z $ is the maximum deviation $ | 1- \sqrt{x} R_Z(x) |_{\rm max} $ of the 
Zolotarev optimal rational polynomial $ R_Z(x) $ of $ 1/\sqrt{x} $ for 
$ x \in [1, \lambda_{max}^2/\lambda_{min}^2] $, with degree $(n-1,n)$ for $ N_s = 2n $.
The optimal weights (\ref{eq:omega}) are used in our 2-flavors simulation,  
with $N_s=2n=16$, $\lambda_{min} = 0.05$ and $\lambda_{max} = 6.2$, 
which gives the maximum deviation $d_Z \simeq 1.1944 \times 10^{-5} $.    

For the simulation of one-flavor, we used the exact one-flavor pseudofermion action 
for domain-wall fermion \cite{Chen:2014hyy}, which requires 
the weights $ \{ \omega_s \} $ satisfying 
the $ R_5 $ symmetry ($ \omega_s = \omega_{N_s -s + 1 } $).
However, (\ref{eq:omega}) does not satisfy the $R_5$ symmetry. 
The optimal $ \{ \omega_s \} $ satisfying $ R_5 $ symmetry 
are obtained in Ref. \cite{Chiu:2015sea}. For $N_s = 2n $, it reads   
\begin{equation}
\label{eq:omega_sym_even}
\omega_s=\omega_{N_s+1-s}=\frac{1}{\lambda_{min}} 
         \sqrt{1-{\kappa'}^2 \mbox{sn}^2 \left( \frac{(2s-1) K'}{N_s}; \kappa' \right)},
\hspace{4mm} s = 1, \cdots, N_s/2,
\end{equation}
where $ \mbox{sn}(u ; \kappa') $ is the Jacobian elliptic function 
with modulus $ \kappa' =\sqrt{ 1 - \lambda_{min}^2 / \lambda_{max}^2 } $,   
and $ K' $ is the complete elliptic function of the first kind with modulus $ \kappa' $.
Then the approximate sign function $ S_{N_s}(H_w) $  
satisfies the bound $ 0 \le 1-S_{N_s}(\lambda) \le 2 d_Z $ 
for $ \lambda^2 \in [\lambda_{min}^2, \lambda_{max}^2] $,  
where $ d_Z $ is defined above. 
Note that in this case $\delta(\lambda)=1-S(\lambda) $ does not satisfy the criterion that the 
maxima and minima of $\delta(\lambda)$ all have the same magnitude but with the opposite sign 
($ \delta_{min} = -\delta_{max} $). However,    
the most salient features of optimal rational approximation of degree $(m,n) $ are preserved, 
namely, the number of alternate maxima and minima is $(m+n+2)$, 
with $ (n+1) $ maxima and $ (m+1) $ minima,
and all maxima are equal to $ 2 d_Z $, while all minima are equal to zero.
This can be regarded as the generalized optimal rational approximation (with a constant shift).  
For our one-flavor simulation, setting $N_s=2n=16$, $\lambda_{min} = 0.05$ and $\lambda_{max} = 6.2$ 
gives the maximum deviation $ 2 d_Z \simeq 2.3889 \times 10^{-5} $.    

%In the following, we give detail of the quark actions in our HMC simulations.

For domain-wall fermions, to simulate $ N_f = 2 +1 + 1 $ 
amounts to simulate $ N_f = 2 + 2 + 1 $ since
\begin{equation}
\label{eq:Nf2p1p1}
\left( \frac{\det \Dodwf(m_{u/d})}{\det \Dodwf(m_{PV})} \right)^2
\frac{\det \Dodwf(m_s)}{\det \Dodwf(m_{PV})}
\frac{\det \Dodwf(m_c)}{\det \Dodwf(m_{PV})} =
\left( \frac{\det \Dodwf(m_{u/d})}{\det \Dodwf(m_{PV})} \right)^2
\left( \frac{\det \Dodwf(m_c)}{\det \Dodwf(m_{PV})} \right)^2
\frac{\det \Dodwf(m_s)}{\det \Dodwf(m_{c})}.
\end{equation}
Obviously, the simulation of 2-flavors with $ (\det \Dodwf(m_c)/\det \Dodwf(m_{PV}))^2 $ 
on the RHS of (\ref{eq:Nf2p1p1}) is more efficient 
than its counterpart of one-flavor with $ \det \Dodwf(m_c)/\det \Dodwf(m_{PV}) $ on the LHS.
Moreover, the one-flavor simulation with $ \det \Dodwf(m_s)/\det \Dodwf(m_{c}) $ on the RHS 
is more efficient than the one with $ \det \Dodwf(m_s)/\det \Dodwf(m_{PV}) $ on the LHS. 
Thus we perform the HMC simulation with the expression on the RHS of Eq. (\ref{eq:Nf2p1p1}). 
%Note that there are 2 different ways to write the $N_f=2+2+1$ expression on the RHS of 
%Eq. (\ref{eq:Nf2p1p1}). However, we have not compared the efficiencies of these 2 different
%options.      

For the two-flavor parts, $ \left(\det \Dodwf(m_{u/d})/\det \Dodwf(m_{PV})\right)^2 $  
and $ \left( \det \Dodwf(m_c)/\det \Dodwf(m_{PV}) \right)^2 $,  
we have implemented two options \cite{Chiu:2011bm,Chen:2019wmn}  
for the $N_f=2$ pseudofermion action in our code, 
and we have used the old action \cite{Chiu:2011bm} in our present simulations. 
Note that both actions give consistent results in the HMC simulations. 
However, if $ \lambda_{min} \leq 0.01 $, 
the new action \cite{Chen:2019wmn} is more efficient than the old one.  
For the old $N_f=2$ pseudofermion action, it can be written as 
\begin{equation}
\label{eq:S_old2F}
S(m_q, m_{PV}) = \phi^\dagger C^\dagger(m_{PV}) \{ C(m_q) C^\dagger(m_q) \}^{-1} C(m_{PV}) \phi, 
\hspace{4mm} m_{PV} = 2 m_0, 
\end{equation}
where
\begin{eqnarray*}
\label{eq:C_def}
C(m_q) &=& 1 - M_5(m_q) D_w^{\text{OE}} M_5(m_q) D_w^{\text{EO}}, \\
\label{eq:M_5}
M_5(m_q) &=& \{ 4 - m_0 + \omega^{-1/2} [1-L(m_q)][(1+L(m_q)]^{-1} \omega^{-1/2} \}^{-1},  
\end{eqnarray*}
and $ L(m_q) $ is defined in (\ref{eq:L}) and (\ref{eq:Lpm}).
Here $\omega \equiv {\rm diag} \{\omega_1, \omega_2, \cdots, \omega_{N_s} \}$ 
is a diagonal matrix in the fifth dimension, and 
$ D_w^{\text{EO}/\text{OE}} $ denotes the part of $ D_w $ with gauge links pointing from 
even/odd sites to odd/even sites after even-odd preconditioning on the 4-dimensional lattice.  
For the $u/d$ quarks, mass-preconditioning \cite{Hasenbusch:2001ne} 
is introduced with two levels of heavy masses: $m_{H_1} \sim 10~m_{u/d} $ and 
$m_{H_2} \sim 100~m_{u/d} $.  
Then the $N_f=2$ pseudofermion action $S(m_{u/d}, m_{PV})$ for $u$ and $d$ quarks 
is replaced with 
\begin{eqnarray*}
\begin{aligned}
S(m_{u/d}, m_{H_1}) + S(m_{H_1}, m_{H_2}) + S(m_{H_2}, m_{PV}) & 
                  \\  
= \phi^{\dagger} C^\dagger(m_{H_1}) \{ C(m_{u/d}) C^\dagger(m_{u/d}) \}^{-1} C(m_{H_1}) \phi 
 + & \phi_1^{\dagger} C^{\dagger}(m_{H_2}) \{C(m_{H_1}) C^\dagger(m_{H_1})\}^{-1} C(m_{H_2}) \phi_1
                  \\  
+ & \phi_2^{\dagger} C^{\dagger}(m_{PV}) \{C(m_{H_2}) C^\dagger(m_{H_2})\}^{-1} C(m_{PV}) \phi_2, 
\end{aligned}
\end{eqnarray*}
which gives the partition function (fermion determinant) 
exactly the same as that of $S(m_{u/d}, m_{PV})$. 

For the one-flavor part, $ \det\Dodwf(m_s)/\det\Dodwf(m_{c}) $, 
we use the exact one-flavor pseudofermion action (EOFA)
for domain-wall fermion \cite{Chen:2014hyy}. 
For optimal DWF, it can be written as ($m_1 < m_2 $) 
\begin{eqnarray}
\label{eq:detDodwf}
\frac{\det\Dodwf(m_1)}{\det\Dodwf(m_2)} 
%= \frac{\det D_T(m_1)}{\det D_T(m_2)}    
= \int d \phi_\pm^\dagger d \phi_\pm \exp\left( - \phi_+^\dagger G_+(m_1,m_2) \phi_+ 
                                   - \phi_-^\dagger G_-(m_1,m_2) \phi_- \right), \hspace{6mm}
\end{eqnarray}
where $ \phi_\pm $ and $ \phi_\pm^\dagger $ are pseudofermion fields 
(each with two spinor components) on the 4-dimensional lattice. Here 
\begin{eqnarray*}
\label{eq:Gp}
G_-(m_1,m_2) &=& P_- \left[I- \left(\frac{m_2 - m_1}{m_2 + m_1}\right) 
                       \omega^{-1/2} v_-^T [H_T(m_1)]^{-1} v_- \omega^{-1/2}\right] P_-, \\ 
G_+(m_1,m_2) &=& P_+ \left[I+ \left(\frac{m_2 - m_1}{m_2 + m_1}\right) 
    \omega^{-1/2} v_+^T [H_T(m_2)-\Delta_+(m_1,m_2) P_+]^{-1} v_+ \omega^{-1/2}\right] P_+,  \\
\label{eq:Gm}
%D_{T}(m_i) &=& D_w + M(m_i), \hspace{4mm} i=1,2\\  
H_T(m_i) &=& R_5 \gamma_5 [ D_w + M(m_i) ], \hspace{4mm} i=1,2\\  
M(m_i) &=& \omega^{-1/2}[1-L(m_i)][1+L(m_i)]^{-1}\omega^{-1/2}, \\ 
%        = P_+ M_+(m_i) + P_- M_-(m_i),  \\ 
%\Delta(m_1,m_2) &=& R_5 \left[ M(m_2)-M(m_1) \right]=P_+ \Delta_+(m_1,m_2)+P_- \Delta_-(m_1,m_1),\\
\Delta_\pm(m_1,m_2) &=& \left(\frac{m_2-m_1}{m_2+m_1}\right)
                        \omega^{-1/2} v_\pm v_\pm^T \omega^{-1/2}, \\ 
%\label{eq:k}
%k(m_1,m_2) &=& \frac{m_2 - m_1}{m_2 + m_1}, \\ 
\label{eq:vp}
v_+^{T} &=& (-1, 1, \cdots, (-1)^{N_s}),  \\
\label{eq:vm}
v_{-} &=& -v_+.
\end{eqnarray*}

\noindent{\bf B. Gauge ensembles} 

In the molecular dynamics, we use the Omelyan integrator \cite{Omelyan:2001abc}
and the multiple-time scale method \cite{Sexton:1992nu}. 
Setting the length of the HMC trajectory equal to one, four different time scales
are used for momentum updates, with the gauge force at level-0,
and the fermion forces at level-1/2/3,  
where the ratio of forces at level-0/1/2/3 is $\sim 1:0.1:0.01:0.001$. 
The step sizes for level-0/1/2/3 are $1/(k_0 k_1 k_2 k_3)$, $1/(k_1 k_2 k_3)$, 
$1/(k_2 k_3)$, and $1/k_3$, where $(k_0, k_1, k_2, k_3)=(10, 2, 2, 12)$ 
is the most common setting in our simulations.  
The momentum updates with the 2-flavor fermion forces corresponding to  
$S(m_{u/d}, m_{H_1})$, $S(m_{H_1}, m_{H_2})$, $ S(m_{H_2}, m_{PV}) $, and $S(m_c, m_{PV})$ 
are set to level-3, levele-2, level-1, and level-1 respectively. 
The momentum updates with the one-flavor fermion forces corresponding to  
$\phi_+ G_+ \phi_+ $ and $\phi_- G_- \phi_- $ are set to level-2 and level-3 respectively.
With the smallest time interval $1/(k_0 k_1 k_2 k_3)$,
the numbers of momentum updates for level-0/1/2/3
are $\{16k_0k_1k_2k_3+1,8k_1k_2k_3+1,4k_2k_3+1,2k_3+1\}$ respectively,
according to the Omelyan integrator.
Our HMC code (DWFQCD) implements the entire HMC trajectory on GPUs, 
in which the most time-consuming parts of computing fermion forces and actions 
are obtained by solving very large and sparse linear systems 
via conjugate gradient with mixed precision. 
%\cite{Chiu:2011rc}.

The initial thermalization of each ensemble was performed in one node  
with 1-8 GPUs interconnected by the NVLink and/or PCIe bus.  
After thermalization, a set of gauge configurations are sampled and distributed
to 8-16 simulation units, and each unit performs an independent stream of HMC simulation.
Here one simulation unit consists of 1-8 GPUs in one node,
depending on the size of the device memory and the computational efficiency.
Then we sample one configuration every 5 trajectories in each stream, and obtain
a total number of configurations for each ensemble.
The statistics of the 15 gauge ensembles with $ T > T_c \sim 150 $~MeV are listed in
Table \ref{tab:15_ensembles_mres}, where $T = 1/(N_t a) $.
For the gauge ensembles with $ T < T_c $, some of them have not reached our desired statistics,  
thus they will be presented in the future.  
For the ensemble of $ 64^4 $ at $ \beta = 6.20 $, 
preliminary results of topological susceptibility and the mass spectra of the low-lying 
mesons and baryons have been presented in Ref. \cite{Chiu:2020ppa}.

\begin{table}[h!]
\begin{center}
\caption{The lattice parameters and statistics of the 15 gauge ensembles with $T > T_c$. 
The last 3 columns are the residual masses of $u/d$, $s$, and $c$ quarks.} 
\setlength{\tabcolsep}{4pt}
\vspace{2mm}
\begin{tabular}{|ccccccccc|}
\hline
    $\beta$
  & $a$[fm]
  & $ N_x $
  & $ N_t $
  & $T$[MeV]
  & $N_{\rm confs}$
  & $ (m_{u/d} a)_{\rm res} $
  & $ (m_{s} a)_{\rm res} $
  & $ (m_{c} a)_{\rm res} $
\\
\hline
\hline
%6.20 & 0.0636 & 64 & 64 & $\sim 0$  &
%$6.33(19) \times 10^{-5} $  & $ 1.62(13) \times 10^{-5} $  &  $ 2.40(39) \times 10^{-6} $  \\
%6.18 & 0.0685 & 64 & 64 & $\sim 0$  &
%$6.93(29) \times 10^{-5} $ & $4.69(25) \times 10^{-6} $  &  $7.64(18) \times 10^{-7} $ \\
%6.15 & 0.0748 & 64 & 64 & $\sim 0$  &
%$1.92(11) \times 10^{-4}$ & $1.50(12) \times 10^{-5} $ & $1.62(27) \times 10^{-6} $     \\
6.20 & 0.0636 & 64 & 20 & 155  & 581 &
$2.39(56) \times 10^{-5}$ & $1.92(53) \times 10^{-5}$ & $7.59(38) \times 10^{-6}$ \\
6.18 & 0.0685 & 64 & 16 & 180 & 650 &
$3.36(32) \times 10^{-5} $ & $1.88(25) \times 10^{-5} $  & $ 5.23(37) \times 10^{-6} $ \\
6.20 & 0.0636 & 64 & 16 & 193 & 1577 &
$1.41(15) \times 10^{-5}$ & $1.14(12) \times 10^{-5}$ & $2.13(28) \times 10^{-6}$ \\
6.15 & 0.0748 & 64 & 12 & 219 & 566 &
$3.16(84) \times 10^{-5} $ &  $2.70(85) \times 10^{-5} $  & $1.24(31) \times 10^{-5} $  \\
6.18 & 0.0685 & 64 & 12 & 240 & 500 &
$2.36(42) \times 10^{-5} $ & $1.72(24) \times 10^{-5} $ & $3.28(57) \times 10^{-6} $  \\
6.20 & 0.0636 & 64 & 12 & 258 & 1373 & 
$2.33(29) \times 10^{-5}$ & $ 2.09(27) \times 10^{-5}$ & $ 6.16(28) \times 10^{-6}$ \\
6.15 & 0.0748 & 64 & 10 & 263 & 690 &
$2.38(36) \times 10^{-5} $  & $1.98(29) \times 10^{-5} $  &  $7.51(26) \times 10^{-6} $  \\
6.18 & 0.0685 & 64 & 10 & 288 & 665 &
$2.42(80) \times 10^{-5} $  & $2.20(73) \times 10^{-5} $  &  $9.74(39) \times 10^{-6} $  \\
6.20 & 0.0636 & 64 & 10 & 310 & 2547 &
$9.61(97) \times 10^{-6}$ & $8.86(96) \times 10^{-6}$ & $2.92(45) \times 10^{-6}$ \\
6.15 & 0.0748 & 64 & 8  & 329 & 1581 & 
$3.24(67) \times 10^{-5}$ & $3.03(62) \times 10^{-5}$ & $1.39(77) \times 10^{-5}$ \\
6.18 & 0.0685 & 64 & 8  & 360 & 1822 &
$2.43(95) \times 10^{-5}$ & $2.24(85) \times 10^{-5}$ & $7.02(25) \times 10^{-6}$ \\
6.20 & 0.0636 & 64 & 8  & 387 & 2665 &
$2.09(86) \times 10^{-5}$ & $1.79(71) \times 10^{-5}$ & $5.72(17) \times 10^{-6}$ \\
6.15 & 0.0748 & 64 & 6  & 438 & 1714 &
$1.61(57) \times 10^{-5}$ & $1.48(50) \times 10^{-5}$ & $8.44(26) \times 10^{-6}$ \\
6.18 & 0.0685 & 64 & 6  & 479 & 1983 &
$8.34(46) \times 10^{-6}$ & $8.26(46) \times 10^{-6}$ & $8.16(49) \times 10^{-6}$ \\
6.20 & 0.0636 & 64 & 6  & 516 & 3038 &
$4.03(82) \times 10^{-6}$ & $3.96(79) \times 10^{-6}$ & $3.06(60) \times 10^{-6}$ \\
\hline
\end{tabular}
\label{tab:15_ensembles_mres}
\end{center}
\end{table}

The lattice spacings and the bare quark masses at the physical point 
are determined on the $64^4$ lattice, with $ \{ 105, 110, 537 \} $ configurations
for $ \beta = \{ 6.15, 6.18, 6.20 \} $ respectively.  
For the determination of the lattice spacing,
we use the Wilson flow \cite{Narayanan:2006rf,Luscher:2010iy} with the condition
\begin{equation*}
\label{eq:t0}
\left. \{ t^2 \langle E(t) \rangle \} \right|_{t=t_0} = 0.3,
\end{equation*}
to obtain $\sqrt{t_0}/a $,
then to use the input $ \sqrt{t_0} = 0.1416(8) $~fm \cite{Bazavov:2015yea} to
obtain the lattice spacing $ a $.
The lattice spacings for $\beta = \{6.15, 6.18, 6.20 \}$ are listed in Table \ref{tab:a_qmass}.
In all cases, the spatial volume satisfies $ L^3 > (4~{\rm fm})^3 $ and $ M_\pi L \gtrsim 3 $.

For each lattice spacing, the bare quark masses of $u/d$, $s$ and $c$ are tuned
such that the lowest-lying masses of the meson operators
$ \{ \bar{u} \gamma_5 d, \bar{s} \gamma_i s, \bar{c} \gamma_i c \} $
are in good agreement with the physical masses of
$\{ \pi^{\pm}(140), \phi(1020), J/\psi(3097) \}$. 
%within error bars $\{5, 8, 10\}$~MeV respectively.
The bare quark masses of $u/d$, $s$, and $c$ of each lattice spacing
are listed in Table \ref{tab:a_qmass}.

\begin{table}[h!]
\begin{center}
\caption{The lattice spacing and the quark masses of the $N_f = 2+1+1$ lattice QCD 
with optimal domain-wall quarks at the physical point.}
\setlength{\tabcolsep}{4pt}
\vspace{2mm}
\begin{tabular}{|ccccc|}
\hline
    $\beta$ 
  & $a$[fm]
  & $ m_{u/d} a $
  & $ m_s a $
  & $ m_c a $ \\
\hline
\hline
6.15 &  0.0748(1) &  0.00200 & 0.064 & 0.705 \\
6.18 &  0.0685(1) &  0.00180 & 0.058 & 0.626 \\
6.20 &  0.0636(1) &  0.00125 & 0.040 & 0.550 \\
\hline
\end{tabular}
\label{tab:a_qmass}
\end{center}
\end{table}  

\noindent{\bf C. Quark propagator}

The valence quark propagator of the 4-dimensional effective Dirac operator can be written as 
$$
(D_c + m_q)^{-1} = \left( 1 - \frac{m_q}{2m_0} \right)^{-1} 
                   \left[ D^{-1}_{N_s}(m_q) - \frac{1}{2m_0} \right]
$$
where $ D_{N_s}(m_q) $ is given in (\ref{eq:Dmq_Ns}), 
and the mass and other parameters are exactly the same as those of the sea quark. 
The boundary conditions of the valence quark propagator 
are periodic in space and antiperiodic in time.
To compute the valence quark propagator, we first solve the following 
linear system with mixed-precision conjugate gradient algorithm,
for the even-odd preconditioned ${\cal D} $ \cite{Chiu:2011rc}
\bea
\label{eq:DY}
{\cal D}(m_q) |Y \rangle = {\cal D}(2m_0) B^{-1} |\mbox{source vector} \rangle,
\eea
where $ B^{-1}_{x,s;x',s'} = \delta_{x,x'}(P_{-}\delta_{s,s'}+P_{+}\delta_{s+1,s'}) $
with periodic boundary conditions in the fifth dimension.
Then the solution of (\ref{eq:DY}) gives the valence quark propagator
\bea
\label{eq:v_quark}
(D_c + m_q)^{-1}_{x,x'} = \left( 2m_0 - m_q \right)^{-1} \left[ (BY)_{x,1;x',1} - \delta_{x,x'} 
\right].
\eea

\noindent{\bf D. Residual mass}

To measure the chiral symmetry breaking due to finite $N_s$ in DWF,  
% and $\lambda_{min} > 0 $, 
we compute the residual mass according to the formula \cite{Chen:2012jya}, 
\begin{equation}
\label{eq:Mres}
(m_q)_{\rm res} = \frac{ \left< \Tr(D_c + m_q)^{-1} \right> }
                { \left< \Tr[\gamma_5 (D_c + m_q)^{-1} \gamma_5 (D_c+m_q)^{-1}] \right>} - m_q,
\end{equation}
where
%$ (D_c + m_q)^{-1} $ denotes the valence quark propagator with $ m_q $ equal to the sea-quark mass,
Tr denotes the trace running over the site, color and Dirac indices,  
and the brackets $ \left< \cdots \right> $ denote the averaging over the configurations 
of the gauge ensemble.
In the limit $ N_s \to \infty $, $ D_c $ is exactly chiral symmetric and the first term on the RHS 
of (\ref{eq:Mres}) is exactly equal to $ m_q $, thus the residual mass $ (m_q)_{\rm res} $ 
is exactly zero, and the quark mass $ m_q $ is well-defined for all gauge configurations. 
However, in practice, $N_s$ is finite, thus the residual mass is nonzero. 
To compute the numerator and the denominator of (\ref{eq:Mres}),  
we use 24-240 $ Z_2 $ noise vectors for each configuration to evaluate 
the all-to-all quark propagators.  
Alternatively, the numerator and the denominator of (\ref{eq:Mres}) can also be estimated with  
the quark propagator from one site (\ref{eq:v_quark}), without summing over all sites. 
It turns out that both methods give consistent results, thus we use their difference     
for the estimate of the systematic uncertainty of the residual mass.   
The residual masses of $u/d$, $s$, and $c$ quarks for the 15 gauge ensembles  
with $ T > T_c $ are listed in the last 3 columns of 
Table \ref{tab:15_ensembles_mres}, 
where the error bar combines both statistical and systematic uncertainties.   
The residual masses of $u/d$, $s$ and $c$ quarks are
less than $1.86\%$, $0.05\%$ and $0.002\%$ of their bare masses respectively. 
In units of MeV/$c^2$, the residual masses
of $u/d$, $s$ and $c$ quarks are less than 0.09, 0.08, and 0.04 respectively.
This asserts that the chiral symmetry is well preserved such that the deviation
of the bare quark mass $m_q$ is sufficiently small in the 
effective 4D Dirac operator $D_{N_s}(m_q)$ of optimal DWF, for both light and heavy quarks.
In other words, the chiral symmetry in our simulations should be sufficiently precise
to guarantee that the hadronic observables
can be determined with a good precision, with the associated uncertainty
much less than those due to statistics and other systematic ones.

\section{Topological charge and Topological susceptibility}
\label{section:chit}

On the lattice, the topological charge $Q_t$ (\ref{eq:Qt}) is ill-defined since we do not have 
$F_{\mu\nu} $ but only link variables. 
To extract $ F_{\mu\nu} $ from the link variables is rather problematic,   
due to the strong short-distance fluctuation. 
The way to circumvent this problem is to   
smooth the link variables with smearing algoithms or the Wilson flow, then  
it is possible to extract $ F_{\mu\nu}(x) $ robustly from the smoothed gauge configuration. 
The resulting $ Q_t $ rounded to the nearest integer serves as a definition of 
the topological charge of this gauge configuration, and the topological susceptibility 
of a gauge ensemble can be measured. 
%This is called the gluonic method.
 
For lattice QCD with exact chiral symmetry, the massless overlap Dirac operator 
in a nontrivial gauge background posseses exact zero modes with definite chirality,  
and its index satisfies the Atiyah-Singer index theorem $ Q_t = n_+ - n_- $, 
where $ n_\pm $ denotes the number of zero modes of $ \pm $ chirality. 
Thus one can project the zero modes of the overlap Dirac operator to obtain the index 
and also the topological charge, without smoothing the gauge configuration at all.
%This is called the fermionic method.
%Note that other lattice Dirac operator do not have exact zero modes 
%with the index satisfying the Atiyah-Singer index theorem.
Nevertheless, it is prohibitively expensive to project the zero modes of the overlap Dirac 
operator for our gauge ensembles with lattice sizes $ 64^3 \times (64, 20, 16, 12, 10, 8, 6) $.  
Thus we use the Wilson flow to measure the topological susceptibility of 
each ensemble. Theoretically, the $ \chi_t(a,T) $ by the Wilson flow  
is not necessarily equal to that using the index of overlap-Dirac operator. 
Nevertheless, both methods should give the same $ \chi(T) $ in the continuum limit. 

\begin{figure}[!ht]
\includegraphics[width=10.0cm,clip=true]{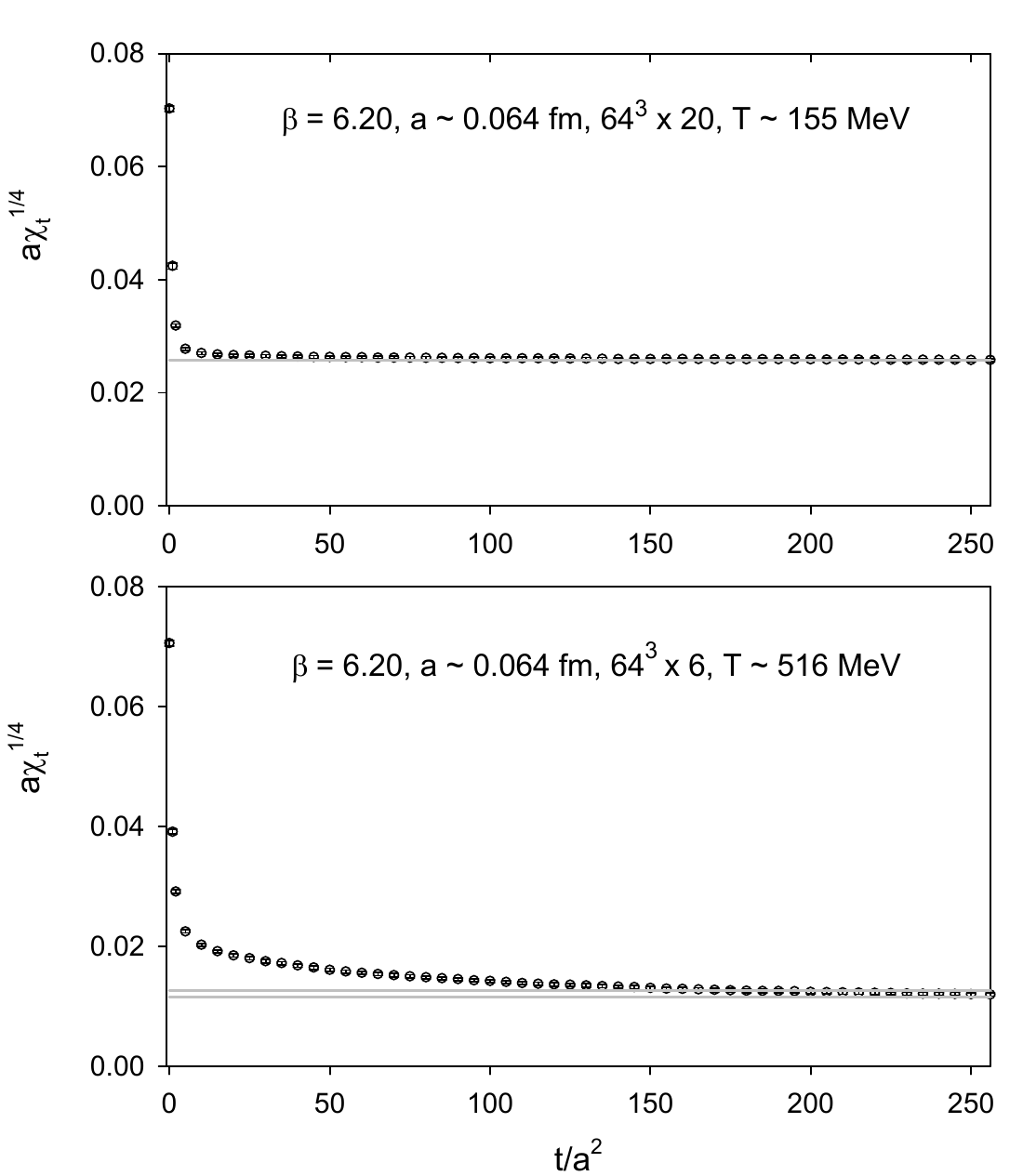} 
\caption{The fourth-root of the toplogical susceptibility $ a \chi_t^{1/4}(T) $ versus  
the flow time $t/a^2$, for $ T \sim 155 $~MeV in the upper panel, and 
$T \sim 516 $~MeV in the lower panel. In each case, the plateau value of $ a \chi_t^{1/4} $ 
is plotted as the horizontal line with the enveloping lines as the error bar. 
}
\label{fig:chit14_b620}
\end{figure}

In this study, the topological charge $Q_t$ of each configuration is
measured by the Wilson flow, using the clover definition.
The Wilson flow equation is integrated from the flow time $t/a^2=0$ to 256
with the step size $0.01$. 
In Fig. \ref{fig:chit14_b620}, the fourth-root of the topological susceptibility 
$ a \chi_t^{1/4}(a,T) $ versus the flow time $t/a^2$ is plotted from $ t/a^2=0 $ to 256, 
for $ T \sim 155 $~MeV (in the upper panel) and $T \sim 516 $~MeV (in the lower panel).
Evidently, as the temperature gets higher, $ \chi_t(a,T) $ attains its plateau value 
at a larger flow time.    

In order to extrapolate the topological susceptibility $ \chi_t = \langle Q_t^2 \rangle/V $
to the continuum limit, $Q_t$ is required to be measured at the same physical flow time
for all lattice spacings, which is chosen to be $0.8192~{\rm fm}^2 $ such that $ \chi_t $
attains its plateau for all gauge ensembles in this study.

\begin{figure}[!ht]
\begin{center}
\includegraphics[width=11cm,clip=true]{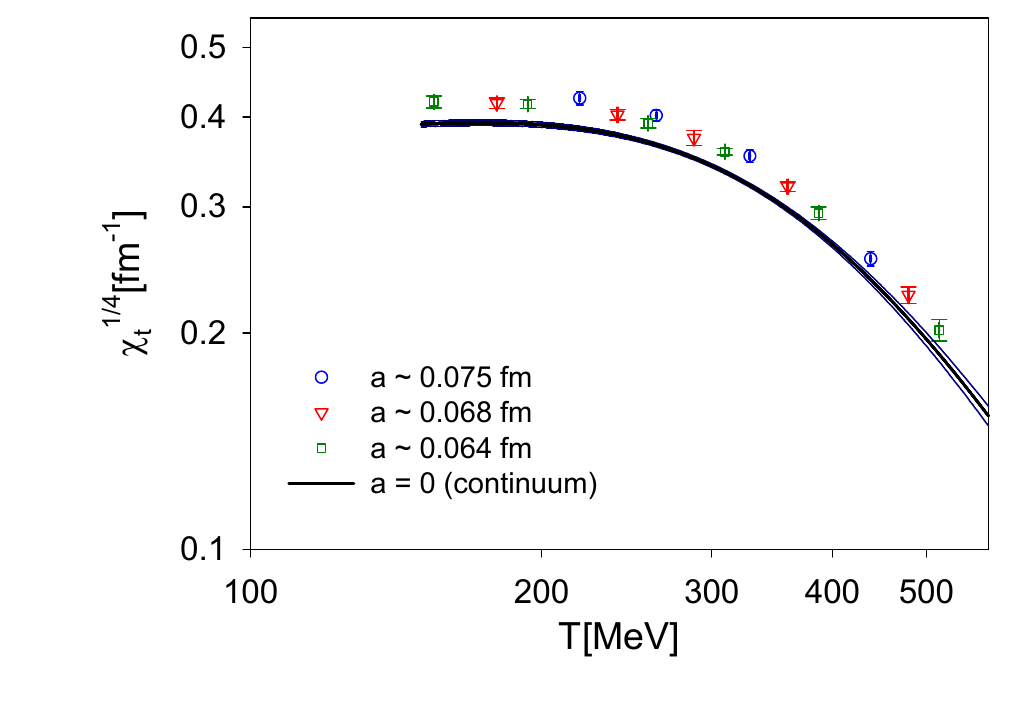}
\caption{
The fourth-root of topological susceptibility $\chi_t^{1/4}(a,T)$ versus the temperature $T$.
The 15 data points with 3 different lattice spacings are denoted by 
blue circles ($a \sim 0.075$~fm), red inverted triangles ($a \sim 0.068$~fm), 
and green squares ($ a \sim 0.064 $~fm). 
The continuum limit resulting from fitting the 15 data points to the ansatz (\ref{eq:ansatz}) is 
denoted by the black line, with the error bars as the enveloping blue lines.}
\label{fig:chit14_15pts_a0}
\end{center}
\end{figure}

The results of the fourth-root of the topological susceptibility 
$\chi_t^{1/4}(a,T)$ (in units of ${\rm fm}^{-1}$) 
of 15 gauge ensembles are listed in the last column of Table \ref{tab:15_ensembles}, 
where the error combines the statistical and the systematic ones. 
Here the systematic error is estimated from the difference of $ \chi_t^{1/4}(a,T) $
using two definitions $ Q_t $, i.e., $ Q_{\rm clover} $ and 
its nearest integer $ {\rm round} [Q_{\rm clover}] $. 
The statistical error is estimated using the jackknife method with the bin size  
of which the statistical error saturates.
The results of $ \chi_t^{1/4}(a,T) $ of 15 gauge ensembles are plotted 
in Fig. \ref{fig:chit14_15pts_a0}.  
They are denoted by blue circles (for $a \sim 0.075$~fm),  
red inverted triangles (for $a \sim 0.068$~fm), and green squares (for $ a \sim 0.064 $~fm).

\begin{table}[h!]
\begin{center}
\caption{The fourth-root of the topological susceptibility $ \chi_t^{1/4}(a,T)$ 
(in units of ${\rm fm}^{-1}$) of the 15 gauge ensembles in this work, 
as a function of the lattice spacing $a$ and the temperature $T$.}
\setlength{\tabcolsep}{4pt}
\vspace{2mm}
\begin{tabular}{|ccccccc|}
\hline
    $\beta$ 
  & $a$[fm]
  & $ N_x $
  & $ N_t $
  & $T$[MeV] 
  & $N_{\rm confs}$ 
  & $\chi_t^{1/4}[{\rm fm}^{-1}]$
\\
\hline
\hline
6.20 & 0.0636 & 64 & 20 & 155 & 545  & 0.420(8) \\ 
6.18 & 0.0685 & 64 & 16 & 180 & 650  & 0.418(7) \\ 
6.20 & 0.0636 & 64 & 16 & 193 & 1577 & 0.417(5) \\ 
6.15 & 0.0748 & 64 & 12 & 219 & 566  & 0.425(9) \\ 
6.18 & 0.0685 & 64 & 12 & 240 & 500  & 0.403(7) \\ 
6.20 & 0.0636 & 64 & 12 & 258 & 1470 & 0.392(6) \\ 
6.15 & 0.0748 & 64 & 10 & 263 & 690  & 0.402(7) \\ 
6.18 & 0.0685 & 64 & 10 & 288 & 665  & 0.374(9) \\ 
6.20 & 0.0636 & 64 & 10 & 310 & 2547 & 0.358(4) \\ 
6.15 & 0.0748 & 64 & 8  & 329 & 1581 & 0.353(7) \\ 
6.18 & 0.0685 & 64 & 8  & 360 & 1822 & 0.320(5) \\ 
6.20 & 0.0636 & 64 & 8  & 387 & 2665 & 0.294(6) \\ 
6.15 & 0.0748 & 64 & 6  & 438 & 1714 & 0.254(6) \\ 
6.18 & 0.0685 & 64 & 6  & 479 & 1983 & 0.226(6) \\ 
6.20 & 0.0636 & 64 & 6  & 516 & 3038 & 0.202(7) \\ 
\hline
\end{tabular}
\label{tab:15_ensembles}
\end{center}
\end{table}  

First, we observe that the 5 data points of $ \chi_t^{1/4}(a,T) $ at high temperature
$ T > 350 $~MeV can be fitted by the power law
$ \chi_t^{1/4}(T) \sim T^{-p} $, independent of the lattice spacing $a$.
However, the power law cannot fit all 15 data points.
In order to construct an analytic formula which can fit all data points of $ \chi_t(T) $
for all temperatures, one considers a function which behaves like
the power law $ \sim (T_c/T)^p $ for $T \gg T_c $, but in general it
incorporates all higher order corrections, i.e.,
\begin{equation}
\label{eq:basic_idea}
\chi_t^{1/4}(T) = c_0 (T_c/T)^{p} \sum_{n=0} b_n (T_c/T)^n. 
%\left[ 1 + b_1 (T_c/T) + b_2 (T_c/T)^2 + \cdot + b_n (T_c/T)^n + \cdots \right].
\end{equation}
In practice, it is vital to recast (\ref{eq:basic_idea})
into a formula with fewer parameters, e.g.,
\begin{equation}
\label{eq:a=0}
\chi_t^{1/4}(T) = c_0 \frac{(T_c/T)^{p}}{1+b_1(T_c/T) + b_2 (T_c/T)^2}.
\end{equation}

It turns out that the 6 data points of $\chi_t^{1/4} $ at $ a \sim 0.064$~fm ($\beta = 6.20 $)
are well fitted by (\ref{eq:a=0}).
Thus, for the global fitting of all $ \chi_t^{1/4}(a,T) $ with different $a$ and $T$,
the simplest extension of (\ref{eq:a=0}) is to replace $ c_0 $ with $ (c_0 + c_1 a^2) $.
This leads to our ansatz
\begin{equation}
\label{eq:ansatz}
\chi_t^{1/4}(a,T)=
(c_0 + c_1 a^2) \frac{(T_c/T)^p}{1 + b_1 (T_c/T) + b_2 (T_c/T)^2}, \hspace{4mm} T_c = 150~{\rm MeV}.
%\chi_t^{1/4}(a,T) = (c_0 + c_1 a^2) \frac{x^p}{1+b_1 x + b_2 x^2},
%                    \hspace{4mm} x \equiv \frac{T_c}{T}, \hspace{2mm} T_c = 150~{\rm MeV},
\end{equation}
Fitting the 15 data points of $ \chi_t^{1/4} $ in Table \ref{tab:15_ensembles}
to (\ref{eq:ansatz}), it gives
$c_0 = 1.89(3)$,
$c_1 = 32.2(6.8)$,
$p = 2.03(5)$,
$b_1 = -2.42(19)$,
$b_2 = 6.25(14)$
with $\chi^2$/d.o.f. = 0.21. 
Note that the fitted value of the exponent $p$ is rather insensitive 
to the choice of $ T_c = 150~{\rm MeV} $,  
i.e., any value of $ T_c $ in the range of 145-155 MeV gives almost the same value of $p$. 
Then $\chi_t^{1/4}(T)$ in the continuum limit can be obtained 
by setting $a^2 =0 $ in (\ref{eq:ansatz}), 
which is plotted as the solid black line in Fig. \ref{fig:chit14_15pts_a0},
with the error bars denoted by the enveloping blue solid lines.
In the limit $ T \gg T_c $, it becomes
$\chi_t^{1/4}(T) = c_0 (T_c/T)^{2.03(5)} $, i.e., $\chi_t(T) = c_0^4 (T_c/T)^{8.1(2)} $,
%\begin{equation}
%\chi_t^{1/4}(T) = c_0 (T_c/T)^{2.03(5)} \iff  \chi_t(T) = c_0^4 (T_c/T)^{8.1(2)},
%\end{equation}
which agrees with the temperature dependence of $\chi_t(T) $ in
the dilute instanton gas approximation (DIGA) \cite{Gross:1980br}, i.e.,
$ \chi_t(T) \sim T^{-8.3} $ for $N_f=4$.
%$ \chi_t(T) \sim T^{-7-N_f/3} $
This also implies that our data points of $ \chi_t(a, T) $ for $ T > 350 $~MeV are valid,
up to an overall constant factor.

It is interesting to note that our 15 data points of $\chi_t(a,T) $
are only up to the temperature $ T \sim 515 $~MeV.
Nevertheless, they are sufficient to fix the coefficents of (\ref{eq:ansatz}),
which in turn can give $ \chi_t(T) $ for any $ T > T_c $.
This is the major advantage of having an analytic formula like (\ref{eq:ansatz}).
There are many possible variations of (\ref{eq:ansatz}), e.g., replacing $(c_0 + c_1 a^2) $
with $ (c_0 + c_1 a^2 + c_2 a^4) $, adding the $a^2$ term to the exponent $ p $
and/or the coefficients $ b_1 $ and $ b_2 $, etc.
For our 15 data points, all variations give consistent results of $ \chi_t(T) $
in the continuum limit.

\section{Volume dependence of the topological susceptibility}
\label{section:volume}

In this Section, we investigate how $\chi_t$ changes with respect to the spatial volume.
To this end, we performed HMC simulations of $N_f=2+1+1$ lattice QCD on the
$ 32^3 \times (64, 16, 12, 10, 8, 6) $ lattices at $ \beta = 6.20 $ with
parameters and $(u/d, s, c)$ quark masses exactly the same as those
in the HMC simulations on the $ 64^3 \times (64, 20, 16, 12, 10, 8, 6) $ lattices.
For each HMC stream after thermalization, we sample one configuration every 5 trajectories
and obtain the total number of configurations of each ensemble. The number of configurations
of each ensemble with $ T > T_c $ is given in the column with header $N_{\rm confs}$
in Table \ref{tab:5_ensembles}.

For the ensemble of lattice size $ 32^3 \times 64 $ at $ \beta = 6.20 $,
the total number of configurations is 187.
Using the Wilson flow and the condition
$ \{ t^2 \langle E(t) \rangle \} |_{t=t_0} = 0.3 $
with the input $ \sqrt{t_0} = 0.1416(8) $~fm \cite{Bazavov:2015yea},
we obtain the lattice spacing $ a = 0.0641(1) $~fm.
We also compute the quark propagators for $u/d$, $s$, and $c$ quarks,
and the time-correlation functions of the meson operators
$ \{ \bar{u} \gamma_5 d, \bar{s} \gamma_i s, \bar{c} \gamma_i c \} $,
and to extract the lowest-lying masses from the time-correlation functions.
The resulting meson masses are in good agreement with the physical masses of
$\pi^{\pm}(140) $, $\phi(1020)$, and $J/\psi(3097)$.

\begin{table}[h!]
\begin{center}
\caption{The fourth-root of the topological susceptibility $ \chi_t^{1/4}(a,T)$
(in units of ${\rm fm}^{-1}$) of 5 gauge ensembles with spatial size $32^3$,
as a function of the temperature $T=1/(N_t a)$.}
\setlength{\tabcolsep}{4pt}
\vspace{2mm}
\begin{tabular}{|ccccccc|}
\hline
    $\beta$
  & $a$[fm]
  & $ N_x $
  & $ N_t $
  & $T$[MeV]
  & $N_{\rm confs}$
  & $\chi_t^{1/4}[{\rm fm}^{-1}]$
\\
\hline
\hline
6.20 & 0.0641 & 32 & 16 & 192 & 1400 & 0.421(12) \\
6.20 & 0.0641 & 32 & 12 & 256 & 755  & 0.398(14) \\
6.20 & 0.0641 & 32 & 10 & 307 & 903  & 0.365(15) \\
6.20 & 0.0641 & 32 & 8  & 384 & 1208 & 0.296(13) \\
6.20 & 0.0641 & 32 & 6  & 512 & 1093 & 0.207(14) \\
\hline
\end{tabular}
\label{tab:5_ensembles}
\end{center}
\end{table}

\begin{figure}[!ht]
\begin{center}
\includegraphics[width=11cm,clip=true]{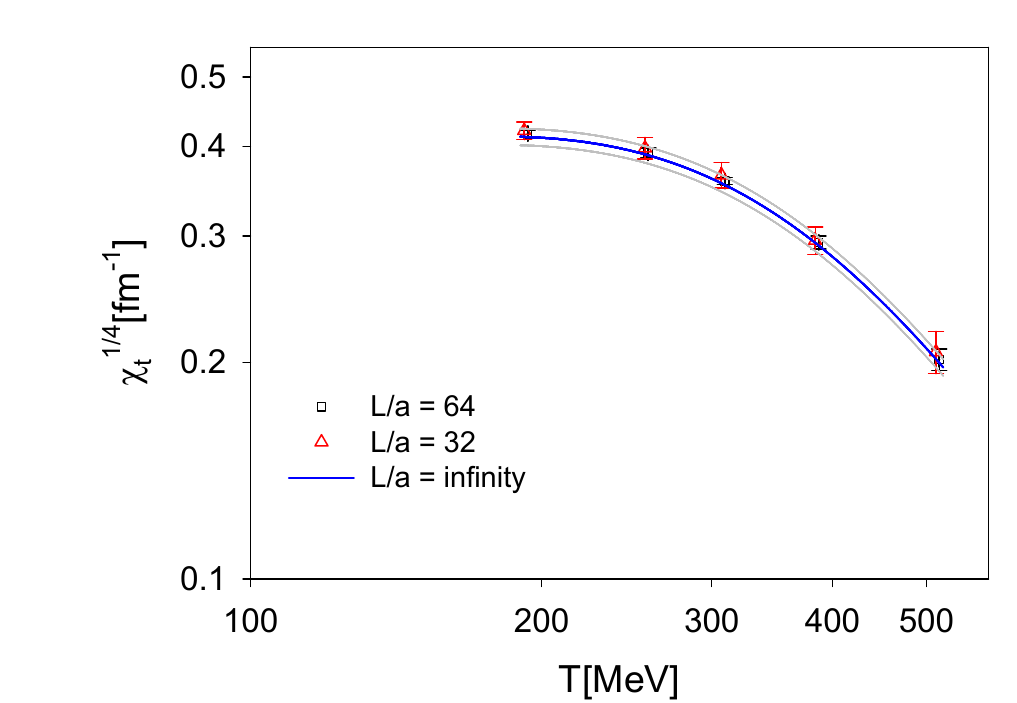}
\caption{
The fourth-root of topological susceptibility $\chi_t^{1/4}(a,T)$ versus the temperature $T$
for $ L/a = 32 $ and $ L/a=64 $ ensembles at $ \beta = 6.20 $.
The 5 data points of the $L/a=32 $ ensembles
are denoted by red triangles, while those of the $L/a = 64 $ ensembles
by black squares.
The infinite volume limit resulting from fitting the 10 data points to
(\ref{eq:ansatz_vol}) is denoted by the blue line,
with the error as the enveloping grey lines.}
\label{fig:chit14_l64_l32}
\end{center}
\end{figure}

Similar to the $L/a=64$ ensembles, the topological charge $Q_t$ of each configuration
in the $ L/a = 32 $ ensembles is measured by the clover definition in the Wilson flow.
The Wilson flow equation is integrated from the flow time $t/a^2=0$ to 256
with the step size $0.01$. The topological charge $Q_t$ of each configuration
is measured at the physical flow time $0.8192~{\rm fm}^2 $,
same as that of any configuration in the $L/a = 64 $ ensembles.
The results of the fourth-root of the topological susceptibility
$\chi_t^{1/4}(a,T)$ (in units of ${\rm fm}^{-1}$)
of these 5 gauge ensembles are listed in the last column of Table \ref{tab:5_ensembles},
where the error combines the statistical and the systematic ones.
Here the systematic error is estimated from the difference of $ \chi_t^{1/4}(a,T) $
using two definitions $ Q_t $, i.e., $ Q_{\rm clover} $ and
its nearest integer $ {\rm round} [Q_{\rm clover}] $.
The statistical error is estimated using the jackknife method with the bin size
of which the statistical error saturates.
Note that due to the lattice spacing $ a = 0.0641(1) $~fm of the $L/a = 32 $ ensembles
is larger than $ a = 0.0636(1) $~fm of the $L/a = 64 $ ensembles,
the temperature $ T = 1/(N_t a) $ of the $ L/a = 32 $ ensemble
in Table \ref{tab:5_ensembles} is slightly lower than that of the $ L/a = 64 $ ensemble
with the same $ N_t $ (see Table \ref{tab:15_ensembles}).
Comparing the results of $\chi_t^{1/4}(a,T)$ in Table \ref{tab:5_ensembles} with
the corresponding ones on the $ 64^3 \times (16,12,10,8,6) $ lattices
in Table \ref{tab:15_ensembles}, we see that
those on the $ 32^3 \times (16,12,10,8,6) $ lattices are all slightly larger than
the corresponding ones on the $ 64^3 \times (16,12,10,8,6) $ lattices, due to
two different volumes as well as two slightly different temperatures.
In Fig. \ref{fig:chit14_l64_l32}, the results of $ \chi_t^{1/4}(a,T) $
of the $ L/a = 32 $ ensembles are plotted as red triangles,
while those of the $ L/a=64 $ ensembles as black squares.
Evidently, the volume dependence of $ \chi_t^{1/4}(a,T) $
for two spatial volumes $ (4.074~{\rm fm})^3 $ and $ (2.053~{\rm fm})^3 $
is smaller than the uncertainty of $\chi_t^{1/4}(a,T)$ of the larger volume,
for all ensembles with $ T > 190 $~MeV.
Thus it is expected that the infinite volume limit ($L/a \to \infty$)
of $ \chi_t^{1/4}(a,T) $ would not be significantly different
from its counterpart on the $ L/a=64 $ lattice.

In general, the spatial volume dependence of $ \chi_t $ (at fixed $ a $ and $T$) can be written as
$$
\chi_t(L) = \chi_t(\infty) \left( 1 + \sum_{n=1}^{\infty} c_n L^{-n} \right). 
$$
In practice, it is reasonable to replace the above expression with
$$
\chi_t(L) = c_0 \exp(c_1/L), \hspace{4mm} c_0 \equiv \chi_t(\infty),
$$
and to determine $ c_0 $ and $ c_1 $ from the data of $ \chi_t(L) $ with different volumes.
Now with 2 sets of $ \chi_t^{1/4}(a,T) $ on two spatial volumes $ (4.074~{\rm fm})^3 $
and $ (2.053~{\rm fm})^3 $, we can extrapolate $ \chi_t^{1/4}(a,T) $ to the infinite volume limit.
Since each set of 5 data points of $\chi_t^{1/4} $ of $ L/a = 32 $ and $ L/a = 64 $ ensembles
is well fitted by Eq. (\ref{eq:a=0}), it is natural to consider the ansatz
\begin{equation}
\label{eq:ansatz_vol}
\chi_t^{1/4}(T,L)= c_0 \exp(c_1/L)
\frac{(T_c/T)^p}{1 + b_1 (T_c/T) + b_2 (T_c/T)^2}, \hspace{4mm} T_c = 150~{\rm MeV}.
\end{equation}
where $ c_0$, $c_1$, $p$, $b$, and $c$ are parameters, and
the dependence on the lattice spacing has been suppressed.
In general, the dependence on the lattice spacing can be incorporated into (\ref{eq:ansatz_vol}),
e.g., replacing $c_0$ with $( d_0 + d_1 a^2 )$.
The infinite volume limit resulting from fitting the 10 data points to
(\ref{eq:ansatz_vol}) is plotted as the blue line, with the error bar as the enveloping grey lines.
Obviously, the $ \chi_t(a,T) $ in the infinite volume limit is in good agreement
with its counterparts of the $ L/a = 64 $ ensembles and the $ L/a = 32 $ ensembles.

For another two sets of $ L/a = 64 $ ensembles at $ \beta = 6.18 $ and $ \beta = 6.15 $,
they have volumes $ (4.384~{\rm fm})^3 $ and $ (4.787~{\rm fm})^3 $, which are larger
than the volume $ (4.074~{\rm fm})^3 $ of the $L/a=64$ ensemble at $ \beta = 6.20 $,
thus it is expected that their finite-volume systematics
are smaller than that of the $ L/a = 64 $ ensemble at $ \beta = 6.20 $.
In other words, for all $ L/a = 64 $ ensembles in this study,
the values of topological susceptibility do not suffer from significant finite-volume systematics.

\section{Comparison with the topological susceptibility by the \\ 
         index of overlap operator} 
\label{section:compare_overlap_clover}

In spite of the fact that our computer resources cannot afford to project 
the zero modes of the overlap operator for any one of the 15 gauge ensembles in this study, 
we can perform the overlap projections for a subset of an ensemble.
Then we can study to what extent the index $(n_+ - n_-)$
of the overlap operator agrees with $ Q_{\rm clover} $ 
in the Wilson flow, and also to compare the $ \chi_t(a,T) $ by the 
overlap index with that by the clover charge in the Wilson flow. 
To this end, we pick the ensemble of $ 64^3 \times 6 $ lattice at $\beta = 6.20$,
with $ a \sim 0.0636 $~fm and $ T \sim 516 $~MeV. 
From 1870 thermalized trajectories generated by the HMC simulation on one unit of Nvidia DGX-V100, 
we sample one configuration every 5 trajectories and obtain 374 configurations 
for the projection of the low modes of overlap Dirac operator \cite{Chiu:2014gna}.
For these 374 configurations, the statistics of the overlap index $ (n_+ - n_-) $ at $ t=0 $
are given in the second column of Table \ref{tab:Q_l64t6_g374}, 
together with those of $ \round[Q_\clover] $
at $ t/a^2 = 25 $ (third column) and $t/a^2 = 256$ (fourth column) respectively.
 
\begin{table}[htb]
\begin{center}
\caption{The statistics of topological charge of 374 configurations 
on the $ 64^3 \times 6 $ lattice at $ T \sim 516 $~MeV.
The second column is the statistics of the overlap index $(n_+ - n_-)$ at $ t=0 $.
The third and the fourth columns are the statistics 
of the nearest integer of clover charge $ \round[Q_\clover]$ at $t/a^2 = 25 $ 
and 256 respectively.}
\setlength{\tabcolsep}{4pt}
\vspace{2mm}
\begin{tabular}{|c|c|cc|}
\hline
$Q_t$ & $ n_+ - n_- $ & \multicolumn{2}{c|}{ ${\rm round}[Q_{\rm clover}]$}  \\
      &               & $t/a^2=25$ & $256$  \\ 
\hline
\hline
-2 &  3   &  0       & \hspace{4mm} 0  \hspace{4mm}  \\
-1 &  48  &  40      & \hspace{4mm} 7 \hspace{4mm}  \\
 0 &  296 &  295     & \hspace{4mm} 356 \hspace{4mm} \\
 1 &  27  &  37      & \hspace{4mm} 11  \hspace{4mm}  \\
 2 &  0   &  2       & \hspace{4mm} 0  \hspace{4mm}  \\
\hline
\end{tabular}
\label{tab:Q_l64t6_g374}
\end{center}
\end{table}

\begin{figure}[!ht]
\includegraphics[width=11.0cm,clip=true]{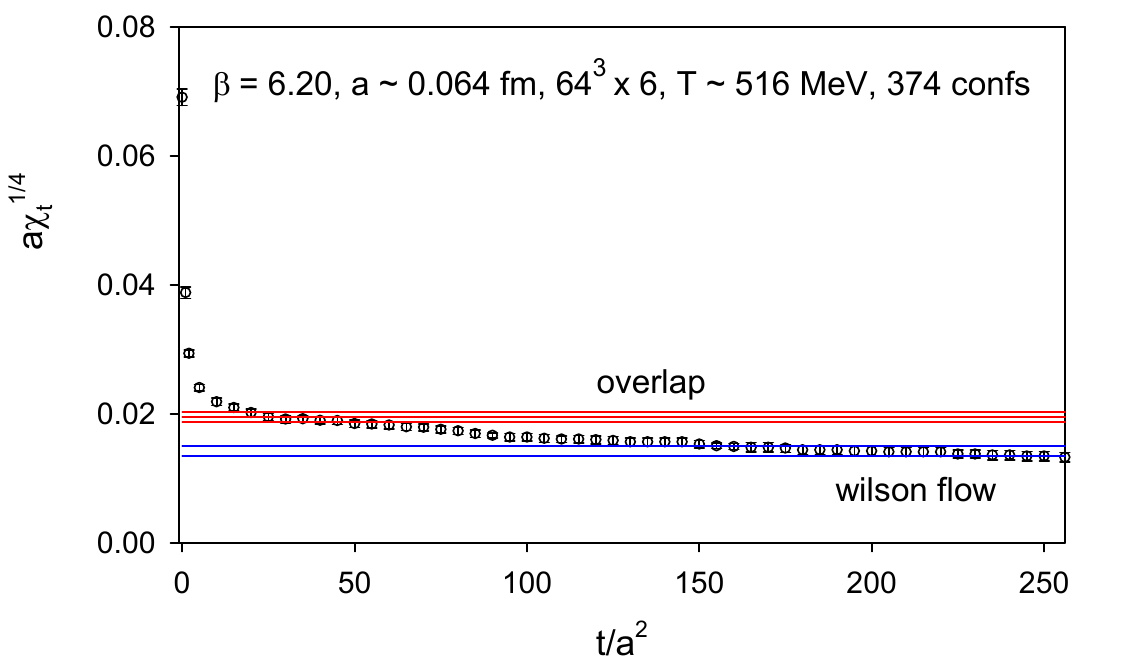} 
\caption{The fourth-root of the toplogical susceptibility $ a \chi_t^{1/4}(a,T) $ of 
374 configurations on the $ 64^3 \times 6 $ lattice at $\beta = 6.20 $, 
with $ a \sim 0.0636 $~fm and $T \sim 516 $~MeV.    
The result of $ a \chi_t^{1/4} $ by the overlap index for the 374 configurations at $t/a^2 = 0 $ is 
denoted by the horizontal red lines, where the center line is the mean value, and the 
upper and lower lines denote the statistical error which is estimated by    
the jackknife method with the bin size at which the statistical error saturates.  
The $ a \chi_t^{1/4} $ by $ \round[Q_\clover] $ in the Wilson flow is  
denoted by circles, and the plateau of $ a \chi_t^{1/4}$ is denoted by the horizontal blue lines. 
}
\label{fig:chit14_l64t6_b620_g374}
\end{figure}

\begin{figure}[!ht]
\begin{center}
\includegraphics[width=11cm,clip=true]{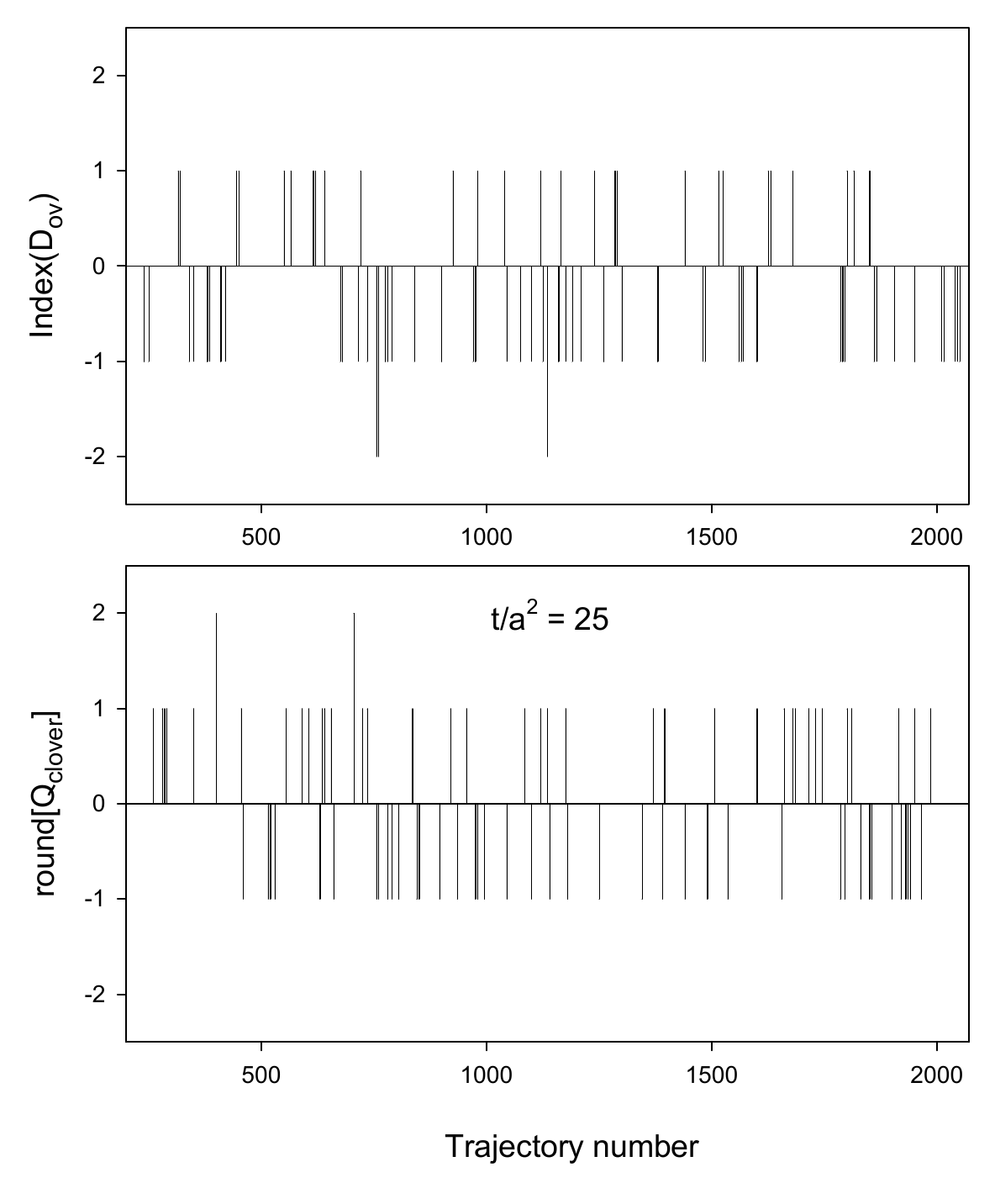}
\caption{The vertical bar plot of the overlap index (in the upper panel) 
and the $ {\rm round}[Q_{\rm clover}] $ at $t/a^2 = 25 $ in the Wilson flow (in the lower panel) 
for the 374 configurations (one configuration every 5 trajectories) 
on the $ 64^3 \times 6 $ lattice at $ T \sim 516 $~MeV.} 
\label{fig:Q_ov_clover_g374}
\end{center}
\end{figure}

Using the the overlap index, the topological susceptibility of these 374 configurations gives 
\bea
\label{eq:chit14_ov}
a \chi_t^{1/4}(a,T)_{\rm overlap} = 0.0195(8),  
\eea 
as shown by the horizontal red lines in Fig. \ref{fig:chit14_l64t6_b620_g374}.
On the other hand, using $ \round[Q_\clover] $ in the Wilson flow, 
the topological susceptibility attains the plateau for $ t/a^2 \gtrsim 180 $, 
and the plateau value gives  
\bea
\label{eq:chit14_clover}
a \chi_t^{1/4}(a,T)_{\rm clover} = 0.0143(8),  
\eea
as shown by the horizontal blue lines in Fig. \ref{fig:chit14_l64t6_b620_g374}.
Theoretically, one does not expect that (\ref{eq:chit14_ov})
could be in good agreement with (\ref{eq:chit14_clover}), 
since at high temperature such as $ T \sim 516 $~MeV, 
the non-trivial topological fluctuations are highly suppressed, 
thus it needs many more than 374 configurations 
in order to obtain reliable statistics for each topological sector. 
Here we recall that in our previous study of $N_f=2$ lattice QCD at zero temperature,  
for an ensemble of 535 configurations on the $24^3 \times 48 $ lattice with $a \sim 0.06~\fm $,     
we find that $ a^4 \chi_t({\rm overlap}) $ at $t=0$ is in
good agreement with the plateau of $ a^4 \chi_t(\clover) $ in the Wilson flow \cite{Chiu:2019xry}. 
Thus we expect that if we could afford to compute the overlap index 
for the entire ensemble of $\sim 3000 $ configurations on the $ 64^3 \times 6 $ lattice 
at $ \beta = 6.20 $, then we may find that  
$ a^4 \chi_t({\rm overlap}) $ at $t=0$ could be in good agreement 
with the plateau of $ a^4 \chi_t(\clover) $ in the Wilson flow.
Most importantly, the values in (\ref{eq:chit14_ov}) and (\ref{eq:chit14_clover}) 
are of the same order of magnitude. 
This seems to justify the plateau value of $ \chi_t(\clover) $      
in the Wilson flow, as well as the results of $ \chi_t^{1/4} $ 
in Table \ref{tab:15_ensembles}. 
       
Note that at $ t/a^2 = 25 $, the topological susceptibility by $ \round[Q_\clover] $ is 
\bea
\left. {a \chi_t^{1/4}(a,T)_{\rm clover}}\right|_{t/a^2 = 25} = 0.0195(5),  
\eea 
which is equal to the $ a \chi_t^{1/4} $ in (\ref{eq:chit14_ov}) by the overlap index at $ t = 0 $. 
However, most of the nontrival configurations according to $ {\rm round}[Q_{\rm clover}] \ne 0 $ 
do not coincide with those according to the overlap index with $ (n_+ - n_-) \ne 0 $ at $ t = 0 $, 
as shown by the vertical bar plot in Fig. \ref{fig:Q_ov_clover_g374}.  
Among the 79 nontrivial configurations with $ {\rm round}[Q_{\rm clover}] \ne 0 $,  
there are only 11 configurations with $ {\rm round}[Q_{\rm clover}] $ equal to the overlap index 
$ (n_+ - n_-) $ at $ t=0 $, 
19 configurations satisfying $ \left| {\rm round}[Q_{\rm clover}] \right| = | n_+ - n_-| $, and 
22 configurations satisfying both $ {\rm round}[Q_{\rm clover}] \ne 0 $ and $ (n_+ - n_-) \ne 0 $. 
In other words, there are 57 nontrivial configurations according to 
$ {\rm round}[Q_{\rm clover}] \ne 0 $, but they are actually trival configurations 
according to the overlap index at $ t = 0 $. Thus, for these 374 gauge configurations, 
even when $ \chi_t $ measured by $ Q_{\rm clover} $ at $ t/a^2 = 25 $ in the Wilson flow 
agrees with that by the overlap index at $ t=0 $, there are $ \sim 72\%$ (57 out of 79) of the 
nontrivial configurations with $ {\rm round}[Q_{\rm clover}] \ne 0 $ 
do not coincide with those with overlap index $ (n_+ - n_-) \ne 0 $. 
Obviously, such a discrepancy becomes larger at other Wilson flow time, where  
$ \chi_t ({\rm clover}) $ is not equal to $ \chi_t ({\rm overlap})$ at $ t=0$.
Moreover, we observe that such a discrepancy commonly exists in any gauge ensemble 
at zero/nonzero temperature. 
 
For completeness, we also project 40 low modes of the overlap Dirac operator $ D_\ov(0) $ with
the Wilson-flowed gauge configuration at $ t/a^2 = 25 $.   
We find that the overlap index is exactly equal to ${\rm round}[Q_\clover] $. 
Then, with the same Wilson-flowed gauge configuration at $ t/a^2 = 25 $,  
we repeat the projection for 40 low modes of 
the effective 4D Dirac operator $ D_{N_s}(0) $ of optimal DWF,  
and find that the eigenvalues of its 40 low modes are almost exactly the same as those 
of the overlap operator $ D_\ov(0) $.
Thus its index is exactly equal to the overlap index and ${\rm round}[Q_\clover] $. 
Then we repeat the same low-mode projections for several Wilson-flowed configurations 
at $ t/a^2 > 25 $, and find that for any one of these configurations,   
the eigenvalues of 40 low-modes of $ D_\ov(0) $ and $ D_{N_s}(0) $ 
are almost exactly the same, and     
$ {\rm index}[D_\ov(0)] = {\rm index}[D_{N_s}(0)] = {\rm round}[Q_\clover] $. 
Thus we conclude that the above equalities must hold 
for any Wilson-flowed configuration at $t/a^2 \ge 25 $.  

Note that the chiral symmetry in our simulation is not exact, with $N_s=16$ and the 
optimal weights $\{ \omega_s \} $ fixed by $ \lambda_{min} = 0.05 $ and $\lambda_{max} = 6.2 $. 
Thus the topological susceptibility in (\ref{eq:chit14_ov}) is obtained by   
a mixed action with $ D_\ov $ in the valence and $ D_{N_s} $ in the sea. 
The most concrete approach to resolve this issue is to perform HMC simulation 
in the exact chiral symmetry limit.        
For optimal DWF, the exact chiral symmetry limit is 
$ N_s \to \infty $ and $ \lambda_{min} \to 0 $. 
This can be attained by increasing $N_s$ and
decreasing $ \lambda_{min} $ such that the systematic error 
due to the chiral symmetry breaking at finite lattice spacing becomes
negligible in any physical observables. For example, if taking
$ N_s=32 $, $\lambda_{min} = 10^{-4} $ and $ \lambda_{max} = 6.2 $, then the error
of the sign function of $H_w$ is less than $1.2 \times 10^{-5} $ for eigenvalues of $ H_w $
satisfying $ 10^{-4} \le | \lambda(H_w) | \le 6.2 $. 
Nevertheless, this set of simulations is estimated to be $ \sim 100$ times 
more expensive than the present one, beyond the limit of our resources. 
In the next section, we examine the feasibility of using the reweighting method 
to correct this systematic error, without new simulations.

\section{The reweighting method}
\label{section:reweighting}

In this section, we discuss the reweighting method to correct 
the systematic error due to the DWF action in the HMC simulation 
not exactly chiral symmetric, without performing new simulations at all. 
Since the reweighting method deforms the path integral non-locally, 
in principle it is not guaranteed to give the correct result.  
Moreover, the reweighting method becomes inefficient if the weights have large fluctuations. 
In the following, we discuss the reweighting method for optimal DWF, which can 
be easily generalized to any DWF. At the end of this section,  
we also discuss the issue of using the clover charge in the Wilson flow
to identify the nontrivial configurations in  
the reweighting method for non-chiral lattice Dirac operators 
(e.g., Wilson Dirac operator, staggered Dirac operator, etc.). 

Consider a set of gauge configurations $\{ U_i, i=1,\cdots, N \} $
obtained by HMC simulation of lattice QCD with $N_f$ domain-wall quarks, of which
the effective 4D Dirac operator is $ D_{N_s}(m_q) $ in (\ref{eq:Dmq_Ns}). 
In the exact chiral symmetry limit, $D_{N_s}(m_q)$ becomes $ D_{\ov}(m_q) $.
The reweighting method to obtain $ \chi_t $ in the exact chiral symmetry limit amounts to compute
\begin{equation}
\label{eq:chit_reweight}
\chi_t^\ov = \frac{1}{V} \frac{\sum_{i=1}^N W_i Q_i^2 }{ \sum_{i=1}^N W_i}
       = \frac{1}{V} \frac{\sum_{Q_i \ne 0} W_i Q_i^2 }{ \sum_{i=1}^N W_i},
\end{equation}
where $ Q_i = {\rm index}\{D_{N_s}(0)[U_i]\} $, and  
\begin{equation}
\label{eq:W}
W_i = \prod_{j=1}^{N_f} \frac{\det D_{\ov}(m_j)[U_i]}{\det D_{N_s}(m_j)[U_i]}. 
\end{equation} 
In general, for any observable ${\cal O} $ 
measured with $\{ U_i \}$ from DWF simulation, its value  
in the exact chiral symmetry limit by the reweighting method is 
\begin{equation}
\label{eq:O_reweight}
\langle {\cal O} \rangle_\ov = \frac{\sum_{i=1}^N W_i {\cal O}_i }{ \sum_{i=1}^N W_i}.
\end{equation}

Note that in (\ref{eq:chit_reweight})-(\ref{eq:O_reweight}),  
$ \{ U_i \} $ are the gauge configurations without any smoothings, 
i.e, at the Wilson flow time $t = 0$. Otherwise, the results in 
(\ref{eq:chit_reweight})-(\ref{eq:O_reweight}) are not well-defined, 
since they depend on how smooth the gauge configurations are. 
For example, consider the ensemble of 374 configurations on the $ 64^3 \times 6 $ lattice as  
described in Section \ref{section:compare_overlap_clover}. 
If one uses the Wilson-flowed gauge ensemble at any $ t/a^2 \ge 25 $ 
for the reweighting in (\ref{eq:chit_reweight})-(\ref{eq:O_reweight}), then 
$ {\rm index}[D_{N_s}(0)] = {\rm index}[D_\ov(0)] = {\rm round}[Q_\clover] $, 
and the weight factor $ W_i = 1 $ for all configurations in the ensemble. 
Consequently, the reweighted $ \chi_t^\ov $ (\ref{eq:chit_reweight}) in the exact chiral symmetry 
limit is the same as that measured by the index of $ D_{N_s}(0) $, or $ {\rm round}[Q_\clover] $, 
i.e., $ \chi_t^\ov = \chi_t^\DWF = \chi_t^\clover $, an incorrect result.        	 
%Thus, for the Wilson-flowed gauge configurations at 
%$ t = 0.8192~{\rm fm}^2 > 25 a^2 $, the above equality is satisfied, 
%and one could {\it incorrectly} claim that 
%the $ \chi_t^{1/4} $ results in Table \ref{tab:15_ensembles} 
%is exactly equal to those in the exact chiral symmetry limit.  
%In the following discussion, (\ref{eq:chit_reweight})-(\ref{eq:O_reweight})  
%always refer to the gauge configurations without any 
%smoothings by the Wilson flow or any smearing algorithms.   

Since $\det D $ is equal to the product of all eigenvalues of $ D $, 
$\det D_{\ov}(m_j)$ can be obtained for any $m_j$ if 
all eigenvalues of $ V_\ov \equiv \gamma_5 H_w/\sqrt{H_w^2} $ are known, i.e., 
from (\ref{eq:Dmq_overlap}),  
\bea
\label{eq:lambda_ov}
\lambda \left[ D_{\ov}(m_j) \right] 
= m_j + \frac{1}{2}( m_{PV} - m_j ) [ 1 + \lambda(V_\ov) ], 
\hspace{4mm} V_\ov \equiv \gamma_5 H_w/\sqrt{H_w^2}. 
\eea
Similarly, $\det D_{N_s}(m_j)$ can be obtained for any $m_j$ if 
all eigenvalues of $ V_{N_s} \equiv \gamma_5 S_{N_s}(H_w) $ are known, 
\bea
\label{eq:lambda_odwf}
\lambda \left[ D_{N_s}(m_j) \right] 
= m_j + \frac{1}{2}( m_{PV} - m_j ) [ 1 + \lambda(V_{N_s}) ], 
\hspace{4mm} V_{N_s} \equiv \gamma_5 S_{N_s}(H_w),  
\eea
where $S_{N_s}(H_w) $ for optimal DWF is defined in (\ref{eq:Dmq_Ns}). 
Note that the counterparts of (\ref{eq:lambda_ov})-(\ref{eq:lambda_odwf}) 
for Shamir/M\"obius DWF can be obtained by replacing $ H_w $ with $ H = \gamma_5 D_w (2+D_w)^{-1} $, 
$ m_{PV} = m_0 (2-m_0) $, and setting $ \{ \omega_s=1, s=1, \cdots, N_s \} $ in $ S_{N_s}(H) $. 
Since $ V_\ov $ is unitary, 
its complex eigenvalues must come in conjugate pairs $\{ e^{i \theta}, e^{-i\theta} \}$, 
with chirality $ \phi^\dagger \gamma_5 \phi = 0 $, where $ \phi $ is the eigenvector. 
Its eigenmodes with real eigenvalues $\pm 1 $ must have chirality $ +1 $ or $ -1 $,  
and satisfy the chirality sum rule \cite{Chiu:1998bh}
\bea
n_+ - n_- + N_+ - N_- = 0,  
\eea
where $ n_\pm $ denotes the number of eigenmodes with eigenvalue $-1$ and chirality $\pm 1 $, 
and $ N_\pm $ denotes the number of eigenmodes with eigenvalue $+1$ and chirality $\pm 1 $. 
Empirically, the real eigenmodes always satisfy either ($ n_+ = N_- $ and $ n_- = N_+ = 0 $)  
or ($ n_- = N_+ $ and $ n_+ = N_- = 0 $). 
According to (\ref{eq:lambda_ov}), 
the $ -1 $ eigenmodes of $V_\ov$ corresponds to the zero mode of $ D_\ov (0) $, 
and the $ +1 $ eigenmodes of $ V_\ov $ corresponds to the nonzero real eigenmodes 
of $ D_\ov(0) $ with eigenvalue $ m_{PV} $, where $m_{PV} = 2 m_0 $ for optimal DWF. 
Thus each zero mode of $ D_\ov(0) $ with definite chirality $ \pm 1 $ must be accompanied with 
a nonzero real eigenmode at $ 2 m_0 $ with opposite chirality $ \mp 1 $. 
For $ V_{N_s} $ in optimal DWF, it is not exactly unitary, but  
it is sufficiently close to unitary such that its eigenvalues are almost the same 
as those of $ V_\ov $ except the real eigenmodes at $\pm 1$. 
In other words, the major difference between the eigenvalues of $D_\ov(0)$ and $D_{N_s}(0) $
are the number of zero modes and the non-zero real eigenmodes at $ 2 m_0 $.   
In the following, we consider all possibilities 
for the zero modes of $D_\ov(0)$ and $D_{N_s}(0)$.  

\begin{enumerate}

\item Both $ D_\ov(0) $ and $ D_{N_s}(0) $ do not have any zero modes. 

In this case, both $ V_\ov $ and $ V_{N_s} $ do not have $ \pm 1 $ real eigenmodes, 
but only complex conjugate pairs which are almost identical for both operators.   
Thus, according to (\ref{eq:lambda_ov})-(\ref{eq:lambda_odwf}), the weight factor (\ref{eq:W}) 
is
\bea
W^{(1)} \simeq 1.
\eea

\item Both $ D_\ov(0) $ and $ D_{N_s}(0) $ have $ n = n_+ + n_- $ zero modes ($ n \ge 1 $). 

In this case, both $ V_\ov $ and $ V_{N_s} $ have $ n $ pairs of real eigenvalues at $+1$ and $-1$.
But the $-1$ eigenvalues of $ V_{N_s} $ could have small deviations,  
say, $-1 + \epsilon_k, k=1, \cdots, n$, where the size of $ \epsilon_k $
depends on how good $ S_{N_s}(H_w) $ can approximate the sign function $ H_w/\sqrt{H_w^2} $, 
especially in the low-lying spectrum of $ | H_w | $, i.e., how small is the $ \lambda_{min} $ 
for computing the weights $\{ \omega_s \} $ in optimal DWF.  
The complex conjugate pairs are almost identical for both $ V_\ov $ and $ V_{N_s} $. 
Then according to (\ref{eq:lambda_ov})-(\ref{eq:lambda_odwf}), 
the weight factor (\ref{eq:W}) for $ N_f=2+1+1$ QCD is 
\bea
\label{eq:W2}
W^{(2)} = 
\prod_{k=1}^n \left[\frac{m_{u/d}}{m_{u/d} +\epsilon_k (m_0-m_{u/d}/2)}\right]^2   
              \left[\frac{m_s}{m_s +\epsilon_k (m_0-m_s/2)} \right]
              \left[\frac{m_c}{m_c +\epsilon_k (m_0-m_c/2)} \right]. \nn
\eea 
Now consider the ensembles at $ \beta = 6.20 $, 
with $ m_{u/d} = 0.00125 $, $ m_s = 0.04 $, $ m_c = 0.55 $ (see Table \ref{tab:a_qmass}), 
and $ m_0 = 1.3 $. 
If a relatively large $ \lambda_{min} $ has been used 
in computing the optimal weights $ \{ \omega_s \} $ for $ V_{N_s} $ 
such that $ \epsilon_k = 0.05 $, then (\ref{eq:W2}) gives 
\bea 
\label{eq:W2ll1}
W^{(2)} \sim \{ (1.888 \times 10^{-2} )^2 \times 0.3846 \times 0.9148 \}^n 
    \sim (1.254 \times 10^{-4} )^n \ll 1, 
\eea 
where $u/d$ quarks at the physical point plays the dominant role in making $ W^{(2)} \ll 1 $.
On the other hand, if a sufficiently small $ \lambda_{min} $ has been used 
in computing the optimal weights $ \{ \omega_s \} $ for $ V_{N_s} $ 
such that $ \epsilon_k \lesssim 10^{-5} $, (\ref{eq:W2}) gives 
\bea
\label{eq:W2sim1}
W^{(2)} \gtrsim \{ (0.98971)^2 \times 0.99968 \times 0.99998 \}^n \sim (0.9792)^n.
\eea 
Thus, in order to make the reweighting method work efficiently, 
$ \lambda_{min} $ is required to be sufficienly small such that $ W^{(2)} \sim 1$.

\item $ D_\ov(0) $ has $n+k$ zero modes ($ n \geq 0 $, $ k \geq 1 $), 
but $ D_{N_s}(0) $ only has $ n $ zero modes. 

First consider the case $ k = 1$. 
Then $ D_\ov(0) $ has one extra zero mode 
plus its accompanying nonzero real eigenmode at $ 2 m_0 $,  
in comparison with the real eigenvalues of $ D_{N_s}(0) $. 
Since the total number of eigenvalues must be the same for both $ D_{\ov}(0) $ and $ D_{N_s}(0) $,  
this implies that $ D_{N_s}(0) $ has one extra complex conjugate pair very close to $2 m_0$, 
in comparison with the complex conjugate pairs of $D_\ov(0) $.    
This can be visualized as follows.
Imagine $ D_{N_s}(0) $ approaching $ D_\ov(0) $ 
by gradually increasing $ N_s $ and decreasing $ \lambda_{min} $, then at some point,     
one of its complex conjugate pair very close to 
$ 2 m_0 $ transform into 2 real eigenmodes, one at zero and the other at $ 2 m_0 $.
The rest of the complex conjugate pairs remain almost identical for both $ D_\ov(0) $ 
and $ D_{N_s}(0) $. Thus the weight factor (\ref{eq:W}) for $N_f = 2+1+1 $ QCD is 
\BAN
\left( \frac{m_{u/d}}{2 m_0} \right)^2 \left(\frac{m_s}{2 m_0} \right) 
\left( \frac{m_c}{2 m_0} \right) W^{(2)}, 
\EAN          
which immediately generalizes to $ k \geq 1 $,  
\bea
\label{eq:W3}
W^{(3)} = \left( \frac{m_{u/d}}{2 m_0} \right)^{2k} \left(\frac{m_s}{2 m_0} \right)^k 
          \left( \frac{m_c}{2 m_0} \right)^k W^{(2)}, 
\eea 
where $ W^{(2)} $ is given in (\ref{eq:W2}). For the ensembles at $ \beta = 6.20 $, 
with quark masses in Table \ref{tab:a_qmass} and $ m_0 = 1.3 $, (\ref{eq:W3}) gives
\BAN
W^{(3)} &\simeq& \left(\frac{0.00125}{2.6} \right)^{2k} 
                 \left(\frac{0.04}{2.6} \right)^k 
                 \left(\frac{0.55}{2.6} \right)^k W^{(2)}
                 \simeq \left(7.522 \times 10^{-10}\right)^k W^{(2)} \ll 1.  
\EAN
If a significant fraction of the nontrivial configurations in the ensemble 
have $W^{(3)} \ll 1 $, then the reweighting method 
cannot work efficiently and the reweighted $ \chi_t $ (\ref{eq:chit_reweight}) is unreliable. 
 
\item $ D_{N_s}(0) $ has $n+k$ zero modes ($ n \geq 0 $, $ k \geq 1 $), 
but $ D_\ov(0) $ only has $ n $ zero modes. 

In principle, this scenario cannot happen  
since $S_{N_s}(H_w) $ is only an approximation of $ H_w/\sqrt{H_w^2} $, 
especially for the low-lying eigenvalues $ 0 < | \lambda(H_w) | < \lambda_{min} $.  
Thus $D_{N_s}(0)$ cannot have more zero modes than $ D_\ov(0) $.

\end{enumerate}

Note that it is rather challenging to compute the weight factor (\ref{eq:W}) numerically, 
since it needs to project the low-lying eigenmodes of both $ V_\ov $ and $V_{N_s}$.
For $ V_\ov $, the projection can be speeded up significantly by low-mode preconditioning
with a few hundred of low-modes of $H_w$ with eigenvalues in the 
range $ 0 < |\lambda(H_w)| \le \lambda_u $, where $ \lambda_u $
depends on the gauge configuration and the number of low-modes.     
Then the sign function $H_w/\sqrt{H_w^2}$ with eigenvalues of $ |H_w| $ 
in the range $ [\lambda_u, 6.2] $      
is approximated by the Zolotarev optimal rational polynomial with 64 poles and 
$ \lambda_{min}/\lambda_{max} = \lambda_{u}/6.2 $. 
On the other hand, for the projection of low-modes of $ V_{N_s} $, 
one is not allowed to use the low modes of $H_w$ for preconditioning.
Otherwise, the corresponding $ D_{N_s}(m_q) $ is not equal to the effective 4D Dirac operator
of the optimal DWF action in the HMC simulation. 
It turns out that the projection of low modes of $ V_{N_s} $ 
is about 5-10 times more expensive than that of $ V_\ov $.  

Testing with the ensemble of 374 configurations on the $64^3 \times 6 $ lattice at $ \beta=6.20 $  
as described in Section \ref{section:compare_overlap_clover},  
we project 40 low modes of $ V_{N_s} $ for the 78 
nontrivial configurations with nonzero overlap index in Table \ref{tab:Q_l64t6_g374}, 
and find that $ V_{N_s} $ does not have any $\pm 1$ real eigenmodes for all of them.   
Thus $ Q_i = {\rm index} \{ D_{N_s}(0) \} = 0 $,  
%and the weight factor $ W \ll 1 $ for these 78 configurations. 
and the reweighted $\chi_t $ (\ref{eq:chit_reweight}) is exactly zero,  
and the reweighting method fails completely in this case.
Next we change $ \lambda_{min} $ from 0.05 to 0.001 and recompute the   
optimal weights $\{\omega_s\}$ for $ S_{N_s}(H_w) $ [see Eq. (\ref{eq:Dmq_Ns})], 
and repeat the low-mode projections.   
Then we find that $ {\rm index} \{ D_{N_s}(0) \} = {\rm index} \{ D_\ov(0) \} = n_+ - n_- $ 
for all 78 nontrivial configurations, but there are $ \sim 20\%$ configurations 
with weight factor $ W < 0.1 $. 
After increasing $N_s$ from 16 to 32 and decreasing $ \lambda_{min} $ from 0.001 to 0.00001, 
then $ {\rm index} \{ D_{N_s}(0) \} = {\rm index} \{ D_\ov(0) \} = n_+ - n_- $,  
and all weight factors are larger than 0.8.
This numerical experiment suggests a viable way to perform the optimal DWF simulation 
such that the resulting gauge configurations are eligible for reweighting to 
the exact chiral symmetry limit. 
That is, to use the optimal weights $\{\omega_s\}$ 
with a sufficiently small $ \lambda_{min} $ and a sufficiently large $ N_s$ 
such that $ V_{N_s} $ has exactly the same number of $ \pm 1 $ real eigenmodes 
as those of $ V_\ov $, and all $-1$ real eigenvalues have 
very small deviations with $ \epsilon_k < 10^{-5} $.   
Then $ {\rm index}[D_{N_s}(0)] = {\rm index}[D_\ov(0)] $  
and the weight factor $W \sim 1 $ for all configurations, and   
the reweighting method works efficiently. 

For completeness, we also project the low modes of $ V_{N_s} = \gamma_5 S_{N_s}(H_w) $ 
with $ S_{N_s}(H_w) $ in polar approximation, which is equivalent to seting $ \omega_s = 1 $
in the optimal DWF. We find that the index of $ D_{N_s}(0) $ with polar approximation 
is zero for all 78 nontrivial configurations with nonzero overlap index 
in Table \ref{tab:Q_l64t6_g374}, for $N_s \le 128 $. 
This implies that the reweighting method also fails 
for other variants (e.g., Shamir/M\"obius) of DWF with $N_s \le 128$. 
Note that for Shamir/M\"obius DWF,  
the approximate sign function is $ S_{N_s}(H) $ with polar approximation, 
where $ H = \gamma_5 D_w (2 + D_w)^{-1} $. 
For any gauge configuration, the low-lying eigenvalues of 
$ H $ should be close to those of $H_w/2$. Thus we expect that 
the $ D_{N_s}(0) $ of Shamir/M\"obius DWF with $ N_s \le 128 $ 
also has zero index for all 78 nontrivial configurations 
with nonzero overlap index in Table \ref{tab:Q_l64t6_g374}, 
even though we have not performed the numerical test. 
This suggests that the viable way to perform Shamir/M\"obius DWF simulation 
such that the resulting gauge configurations are eligible for reweighting to 
the exact chiral symmetry limit is to use a sufficiently large $ N_s $ which 
is much larger than that of optimal DWF, since it does not have 
any parameter like $ \lambda_{min} $ to enhance the chiral symmetry 
for the low-lying eigenvalues of $|H|$. 

In the following, we discuss the issue of using the clover charge in the Wilson flow
to identify the nontrivial configurations in the reweighting method (\ref{eq:chit_reweight}). 
%(with configurations at Wilson flow time $t=0$) 
For non-chiral lattice Dirac operators 
(e.g, the Wilson Dirac operator, the staggered Dirac operator, etc.), 
they do not have exact zero modes at finite lattice spacing.  
Thus, it is impossible to use the eigenvalues of any non-chiral lattice Dirac operator 
to identify the topologically nontrivial configurations in (\ref{eq:chit_reweight}).     
If the clover charge in the Wilson flow is used to identify the nontrivial configurations, 
it could be different from the index of the 
non-chiral lattice Dirac operator in the continuum limit. 
As demonstrated in Section \ref{section:compare_overlap_clover},  
for an ensemble of 374 gauge configurations at $ T \sim 516 $~MeV, even
at the Wilson flow time $ t/a^2 = 25 $ where $ \chi_t $ measured by $ {\rm round}[Q_{\rm clover}] $  
is almost equal to that by the overlap index at $ t=0 $, 
there are more than $72\%$ (57 out of 79) 
of the configurations with $ {\rm round}[Q_{\rm clover}] \ne 0 $ 
but with the overlap index $ (n_+ - n_-) = 0 $ at $t=0$.
For a crosscheck, one can use the index of overlap operator to identify
the nontrivial configuratons for reweighting, to check whether 
the reweighted $\chi_t$ in the continuum limit is consistent with that obtained by 
using the clover charge in the Wilson flow to identify the nontrivial configurations.

\section{Comparison with other lattice studies}
\label{section:compare_chit}

\begin{figure}[H]
\centering
\includegraphics[width=11cm,clip=true]{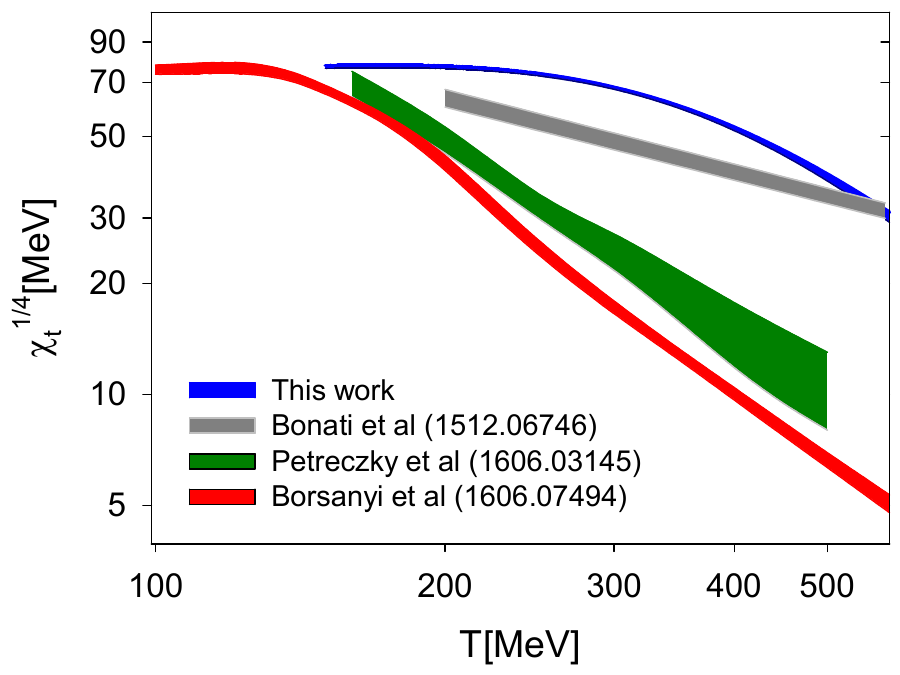}
\caption{
Comparsion of the continuum extrapolated 
fourth-root topological susceptibility $\chi_t^{1/4}(T)$ for four lattice studies.}
\label{fig:chit14_comparison}
\end{figure}

In the following, we survey the continuum extrapolated $ \chi_t(T) $ in recent lattice studies 
with $N_f = 2+1(+1) $ dynamical fermions at/near the physical point, 
and discuss their discrepancies.     
In Fig. \ref{fig:chit14_comparison}, results of four lattice studies are plotted,    
while other results not shown are either not in continuum limit or only a single data point 
at one temperature, and they will be included in the following discussions.       
Note that the data points of Bonati et al. \cite{Bonati:2015vqz} 
and Petreczky et al. \cite{Petreczky:2016vrs} are read off from the figures in the original  
publications, thus they may have large uncertainties due to the limited resolution of human eyes.

The results of Bonati et al. \cite{Bonati:2015vqz} 
were obtained from direct simulations of $N_f = 2+1$ lattice QCD at the physical point 
(with $m_\pi = 135$~MeV, and $ m_s/m_{ud} = 28.15 $), 
using the tree level improved Symanzik gauge action and the stout improved staggered fermion action. 
The continuum limit of $\chi^{1/4} $ was obtained by extrapolation with 
three lattice spacings $ a = (0.0572, 0.0707, 0.0824)$~fm. 
The topological charge of each configuration was measured by the clover charge after cooling.   

The results of Petreczky et al. \cite{Petreczky:2016vrs} were obtained from simulations  
of $ N_f = 2+1$ lattice QCD with $m_\pi = 160$~MeV and $m_s/m_{ud} = 20 $ (physical $m_s$), 
using the tree-level improved gauge action and the highly improved staggered quark action (HISQ).  
The continuum limit of $\chi_t^{1/4} $ is obtained by extrapolation with three 
lattice spacings with $N_t=(8,10,12)$. 
They used two methods to measure the topological susceptibility:    
(1) The clover charge in the Symanzik flow;  
(2) The chiral susceptibilities $ \chi_\pi $ and $ \chi_\delta $ 
and the relation $ \chi_t = m_{ud}^2 \chi_{\rm disc} $ for $ T > T_c $; 
and both methods gave compatible results.   
In Fig. \ref{fig:chit14_comparison}, 
only the data points obtained with the clover charge are plotted.   

The topological susceptibility of Borsanyi et al. \cite{Borsanyi:2016ksw}    
was measured by the clover charge in the Wilson flow, and the data points in  
Fig. \ref{fig:chit14_comparison} are based on the numerical results 
%of $ \log_{10}(\chi[{\rm fm}^{-4}]) $ 
in Table S9 of the  
\href{https://static-content.springer.com/esm/art%3A10.1038%2Fnature20115/MediaObjects/41586_2016_BFnature20115_MOESM349_ESM.pdf}{Supplementary Information} 
of Ref. \cite{Borsanyi:2016ksw},    
which are supposed to be the continuum extrapolated topological susceptibility 
of $ N_f = 2+1+1 $ QCD at the physical point, plus the theoretically estimated 
contribution of the $b$ quark and the correction for the mass difference 
between $u$ and $d$ quarks. 
However, only seven data points in the range of $ T = 130-300 $~MeV 
were based on direct simulations of $N_f = 2 + 1 + 1 $ lattice QCD at the physical point, 
using the tree-level Symanzik gauge action 
and the staggered quark action with 4 levels of stout smearing. 
For other data points, they were obtained by the fixed sector intergal 
and the eigenvalue reweighting techniques from three sets of unphysical simulations: \\
(a) $N_f=3+1$ (three flavors of physical $m_s$ and one flavor of physical $m_c$) 
    for $ T = 150-500 $~MeV;\\ 
(b) Same as (a) but at fixed topology for $ T=300-3000$~MeV; \\
(c) $N_f=2+1$ overlap fermions at fixed topology for three temperatures, $T=(300,450,650)$~MeV, 
    and each for 6 $m_{ud}$ quark masses between physical $m_s$ and physical $m_{ud}^{\rm phys}$. \\ 
Thus, for comparison with other lattice results, we focus on their data points 
in the range of $T=150-300$~MeV, which were obtained by direct simulations at the physical point, 
corrected by the eigenvalue reweighting, and extrapolated to the continuum limit. 
 
First, we compare the results of Bonati et al. \cite{Bonati:2015vqz}, 
Petreczky et al. \cite{Petreczky:2016vrs} and Borsanyi et al. \cite{Borsanyi:2016ksw}.  
Evidently, the discrepancies between Petreczky et al. and Borsanyi et al. are much smaller 
than those between Bonati et al. and Borsanyi et al.  
Moreover, after the results of Petreczky et al. \cite{Petreczky:2016vrs} 
are transformed from $m_\pi = 160 $~MeV to the physical point by the 
relation $ \chi_t^{1/4} \propto m_\pi $, they seem to be in good agreement with      
the results of Borsanyi et al. \cite{Borsanyi:2016ksw}. 

In a more recent study by Bonati et al. \cite{Bonati:2018blm}   
in $N_f=2+1$ lattice QCD at the physical point with
tree level improved Symanzik gauge action and the stout improved staggered fermion action,  
using the multicanonical algorithm (to enhance the topological fluctuations),  
they obtained the continuum extrapolated $ \chi_t^{1/4} = (3 \pm 3 \pm 2) $~MeV 
at $T \simeq 430$~MeV, which is $\sim 9 \sigma $ 
different from their previous result $ \sim 38(2) $~MeV in Ref. \cite{Bonati:2015vqz}. 
The topological charge of each configuration is measured by the clover charge after cooling.   
Then, in the most recent study of the same group \cite{Athenodorou:2021rtv},  
using the same set of ensembles at $T \simeq 430 $~MeV \cite{Bonati:2018blm}, 
they obtained the continuum extrapolated $\chi_t^{1/4} \sim 20 (3) $~MeV 
(read off from Fig. 2 of Ref. \cite{Athenodorou:2021rtv}), 
which is $ \sim 5 \sigma $ different from their 2018 result \cite{Bonati:2018blm}, 
and $\sim 3 \sigma $ different from $9(1)$~MeV of Borsanyi et al. \cite{Borsanyi:2016ksw}.    
Note that in Ref. \cite{Athenodorou:2021rtv}, two methods had been used to measure the $\chi_t$:    
(1) The index of the staggered spectral projector; 
(2) The clover charge after cooling;
and both methods gave compatible results. 
   
In Table \ref{tab:comparison}, we compile all results of continuum extrapolated 
$\chi_t^{1/4} $ at $T \simeq 430 $~MeV, 
together with their lattice actions, simulation methods and techniques, 
and methods (gluonic and fermionic ones) for $\chi_t$ measurement.    
     
\begin{table}[h!]
\begin{center}
\caption{The continuum extrapolated 
fourth-root of the topological susceptibility $ \chi_t^{1/4}[{\rm MeV}]$ 
at $T \simeq 430$~MeV. The abbreviations are:
DWF (Domain-wall fermion), 
HISQ (Highly improved staggered quark),   
SISF (Stout-improved staggered fermion), 
WTMF (Wilson twisted mass fermion);  
Wilson (Wilson plaquette action), 
Symanzik (Tree-level improved Symanzik gauge action),  
Iwasaki (Iwasaki gauge action with $c_0 = 3.648$ and $c_1=-0.331$); 
REW (Eigenvalue reweighting technique), 
FSI (Fixed sector integral technique),  
MCA (Multicanonical algorithm);
CCC  (Clover charge after cooling), 
CSF  (Clover charge in the Symanzik flow), 
CWF  (Clover charge in the Wilson flow), 
DIS  (Use $ \chi_t = m_{ud}^2 \chi_{\rm disc} $), 
SSP  (Staggered spectral projectors).
}
\setlength{\tabcolsep}{4pt}
\vspace{2mm}
\begin{tabular}{|cccccccc|}
\hline
    Reference
  & Quark  
  & Gluon
  & $N_f$
  & $m_\pi[{\rm MeV}]$
  & Simulation
  & Measurement
  & $\chi_t^{1/4}[{\rm MeV}]$ 
\\
\hline
\hline
Bonati et al.\cite{Bonati:2015vqz} & SISF & Symanzik & 2+1 & 135 & Direct & CCC & $ 38(2)$ \\
Petreczky et al. \cite{Petreczky:2016vrs} & HISQ & Symanzik & 2+1 & 160 & Direct & CSF, DIS & 15(2), 10(3) \\ 
Borsanyi et al. \cite{Borsanyi:2016ksw} & SISF & Symanzik &  2+1+1 & 140 & REW, FSI & CWF & 9(1) \\  
Bonati et al. \cite{Bonati:2018blm} & SISF  & Symanzik & 2+1 & 140 & Direct+MCA  & CCC & 3(3)(2) \\ 
Athenodorou et al. \cite{Athenodorou:2021rtv} & SISF & Symanzik & 2+1 & 140 & Direct+MCA  & CCC, SSP & 18(3), 20(3) \\ 
Kotov et al. \cite{Kotov:2021ujj} & WTMF & Iwasaki & 2+1+1 & 140 & Direct & DIS & $ < 4(1) $  \\ 
This work & DWF & Wilson & 2+1+1 & 140 & Direct & CWF & 48(1) \\
\hline
\end{tabular}
\label{tab:comparison}
\end{center}
\end{table}

We note that there are ongoing studies of $\chi_t(T)$ in $N_f=2+1+1$ lattice QCD 
with Wilson twisted mass fermions \cite{Burger:2018fvb,Kotov:2021ujj}.  
Using the relation $ \chi_t = m_{ud}^2 \chi_{\rm disc} $ to measure $ \chi_t $ 
via the noise estimation of the disconnected chiral susceptibility of $u/d$ quarks, 
they obtained $\chi_t^{1/4} \sim 10(2) $~MeV at $T\simeq 430$~MeV,  
with $ m_\pi = 210$~MeV and $a \sim 0.065$~fm \cite{Burger:2018fvb}. 
Their recent results at the physical point with $ m_\pi = 139(1)$~MeV 
and $a \sim 0.080$~fm were presented in Ref. \cite{Kotov:2021ujj},    
and at $T \simeq 430$~MeV, $\chi_t^{1/4} \sim 4(1) $~MeV  
(read off from Fig. 2 of Ref. \cite{Kotov:2021ujj}).   
This implies that the continuum extrapolated $\chi^{1/4} $ at $T \simeq 430$~MeV 
would be less than $4(1)$~MeV.  
This is added to Table \ref{tab:comparison} for comparison with 
other continuum extrapolated $\chi_t^{1/4}$ at the same temperature. 
%
%If the cutoff effects due to the non-chiral Wilson twisted mass fermions are also corrected,  
%its value would be even smaller. 
%Also it is interesting to check whether the gluonic method would give
%$\chi_t$ in agreement with their present results by the fermionic method.  

The discrepancies of the continuum extrapolated $\chi_t^{1/4}$ 
shown in Fig. \ref{fig:chit14_comparison} and in the last column of Table \ref{tab:comparison} 
suggest that the systematic errors in all/most lattice studies have not been under control. 
Note that, except Borsanyi et al. \cite{Borsanyi:2016ksw},     
all lattice results have not been corrected for the cutoff effects
due to the lattice Dirac operator in a nontrivial gauge background 
not possessing (or not having the complete set of) exact zero modes. 
%
% Newly added on August 3, 2022
%
That is, such cutoff effects of the lattice Dirac operator were corrected 
at finite lattice spacing before extrapolating $\chi_t(a,T)$ to the continuum limit.
Otherwise, the results of $\chi_t(T) $ in the continuum will be suffered from 
large cutoff effects. 
Now the question is what would be the scenario 
if all lattice results are corrected for these cutoff effects.
Since there were four studies using the stout-improved staggered fermions 
and one of them had already performed the reweighting \cite{Borsanyi:2016ksw},     
we can use the effective reweighting factors obtained in Ref. \cite{Borsanyi:2016ksw} 
to get a rough estimate of the reweighted $\chi_t^{1/4}$ in another three studies    
\cite{Bonati:2015vqz, Bonati:2018blm, Athenodorou:2021rtv}. 

According to the data in Table S8 and Fig. 25 of the  
\href{https://static-content.springer.com/esm/art%3A10.1038%2Fnature20115/MediaObjects/41586_2016_BFnature20115_MOESM349_ESM.pdf}{Supplementary Information} 
and the Extended Data Fig. 4 in Ref. \cite{Borsanyi:2016ksw},  
the reweighting at $T=300$~MeV effectively imposes a factor $\sim 0.38$ 
to the continuum extrapolated $\chi_t^{1/4}$.
Note that the reweighting factor for lattice QCD with 
$N_f=2+1+1$ staggered fermions  
is almost the same as that for lattice QCD with $N_f=2+1$ staggered fermions, 
since the mass of the charm quark is much larger than   
the eigenvalue of the would-be zero mode of the massless staggered fermion operator. 
That is, the reweighting factor of the staggered charm quark is  
$$ 
\prod_{i=1}^{2|Q_t|} \left( \frac{m_c^2}{m_c^2 + |\lambda_i|^2} \right)^{1/4} \simeq 1,  
\hspace{4mm}  m_c \gg |\lambda_i|. 
$$
Thus the same effective reweighting factor $\sim 0.38$ can be used 
for a rough estimate of the reweighted $\chi_t^{1/4}$ at $T=300$~MeV 
in another $N_f=2+1$ lattice QCD study with stout-improved staggered fermions. 
In the case of Ref. \cite{Bonati:2015vqz}, it changes the value of $\chi_t^{1/4}$ 
from $\sim 50(2)$~MeV to $\sim 19(2)$~MeV, bringing it
into good agreement with the value $17(1)$~MeV of Borsanyi et al. \cite{Borsanyi:2016ksw}.  

Next, we turn to the results of $ \chi_t^{1/4} $ in Table \ref{tab:comparison}. 
However, for $T=430$~MeV, the effective reweighting factor 
for $N_f=2+1+1$ stout-improved staggered fermions 
is not available in Ref. \cite{Borsanyi:2016ksw}, 
since the simulation at $T=430$~MeV was only performed for unphysical $N_f=3+1$ lattice QCD. 
Nevertheless, it must be smaller than the value $0.38$ at $T=300$~MeV,  
since the eigenvalue of the would-be zero mode becomes larger at higher $T$. 
For a very rough estimate, we take it to be $0.35$ at $T=430$~MeV, 
and apply it to the entries of \cite{Bonati:2015vqz, Bonati:2018blm, Athenodorou:2021rtv}  
in the last column of Table \ref{tab:comparison}.
This gives the ``reweighted" $\chi_t^{1/4} $: 
$ 13(1) $~MeV for \cite{Bonati:2015vqz}; 
$ 2(2)(1) $~MeV for \cite{Bonati:2018blm};   
and $ \{ 6(2), 7(2) \}$~MeV for \cite{Athenodorou:2021rtv}. 
Thus, after such a reweighting at $T\simeq 430$~MeV, 
Bonati et al. \cite{Bonati:2015vqz} and Athenodorou et al. \cite{Athenodorou:2021rtv} 
become in closer agreement with Borsanyi et al. \cite{Borsanyi:2016ksw}, 
while Bonati et al. \cite{Bonati:2018blm} in larger disagreement with 
Borsanyi et al. \cite{Borsanyi:2016ksw}. 

For a lattice fermion operator different from the 
stout-improved staggered fermion operator,  
the effective reweighting factor obtained in Ref. \cite{Borsanyi:2016ksw} 
cannot be used for a rough estimate of the reweighted $\chi_t^{1/4}$, 
since their cutoff effects could be very different. 
In general, for any lattice Dirac operator (except the overlap operator), 
the reweighted $\chi_t(T)$ would be smaller than that without reweighting, 
and the reduction becomes larger at higher $T$. 
This suggests that the continuum extrapolated $\chi_t(T)$ of all lattice studies 
could be brought into agreement if the cutoff effects of the lattice Dirac operators 
would be corrected.   

Recall that the reweighting method deforms the path integral non-locally, 
in principle it is not guaranteed to give the correct result. 
Moreover, the reweighting method becomes inefficient if the weights have large fluctuations. 
Thus it is necessary to crosscheck the results of Borsanyi et al. \cite{Borsanyi:2016ksw} 
by direct simulations with overlap fermions.

For domain-wall fermions (DWF), the viable way to reduce the cutoff effects  
is to perform simulations with more and more precise chiral symmetry successively. 
On the other hand, if one resorts to reweighting, the prerequisite for a reliable reweighting     
is that the chiral symmetry of the DWF should be sufficiently precise 
such that for any nontrivial gauge background its effective 4D Dirac operator 
possesses the same number of exact zero modes as the overlap operator, 
as discussed in Section \ref{section:reweighting}. 

About the results of continuum extrapolated $\chi_t^{1/4}(T)$ in this work, 
they are the largest among all lattice results at the same temperature. 
Theoretically, they would become smaller in the exact chiral symmetry limit 
(with $ N_s \to \infty $ and $ \lambda_{min} \to 0 $). 
At this moment, we do not know to what extent the decrease would be, and 
whether they would agree with the results of Borsanyi et al. \cite{Borsanyi:2016ksw}.
Note that we have used the Wilson plaquette action and  
the optimal DWF operator with thin links, unlike other lattice 
studies using improved gauge actions and the stout-improved lattice Dirac operator.
Thus the topological fluctuations (and the susceptibility) in our case 
are expected to be larger than those using other actions 
at the same temperature and lattice spacing.  
It is unclear to what extent the cutoff effects due to the gauge action and 
the link variables entering the lattice Dirac operator 
could affect the continuum extrapolated $\chi_t(T)$, even though 
in principle all cutoff effects are supposed to vanish in the continuum limit. 
On the other hand, it is also unclear whether using improved gauge actions 
and improved lattice Dirac operators with fat links would suppress the topological fluctuations 
too much such that the continuum extrapolation would be distorted. 
Further studies are required to answer these questions.   

Other systematics in Ref. \cite{Borsanyi:2016ksw} 
are the fixed-sector integral method (in $T$) together with the reweighting method 
to extend the unphysical $N_f=3+1$ simulations from 500 MeV to 3000 MeV, 
and also the integral method (in $m_{ud}$) to bring the 
unphysical $N_f=3+1$ results from $m_{ud} = m_s^{\rm phys}$ to physical $m_{ud}^{\rm phys}$.     
These systematics can be crosschecked by direct simulations with $N_f=2+1+1$ overlap fermions
at the physical point and without fixing topology.
%at least for one temperature, 
%say $T \simeq 430$~MeV. 
Since direct simulations of overlap fermions at the physical point is 
prohibitively expensive, a viable alternative is to use the optimal DWF with 
sufficiently small $\lambda_{min}$ and sufficiently large $N_s$, which may be 
feasible with the exaflop machines. 

To conclude this Section, we reiterate that the systematic errors of all/most lattice 
results of $\chi_t(T)$ have not been under control, leading to the discrepancies 
as shown in Fig. \ref{fig:chit14_comparison} and in Table \ref{tab:comparison}. 
Moreover, any convergence of several lattice results at some temperature 
does not necessarily implies that it would be the correct physical/theoretical value, 
which can be established only after the systematic errors of all these lattice results 
have been corrected.

%\vfill
\section{Summary and concluding remarks}
\label{section:summary}

To determine the topological susceptibility of finite temperature QCD is a very challenging task.
So far all lattice studies have not obtained satisfactory results with all systematic errors 
under control, at the physical point as well as in the continuum limit, for $ T > T_c $. 

The present study is the first attempt to simulate finite temperature lattice QCD 
with $N_f = 2+1+1 $ domain-wall quarks at the physical point.
We perform the HMC simulation of lattice QCD with $N_f=2+1+1$ 
optimal domain-wall quarks at the physical point,
on the $64^3 \times (64, 20, 16, 12, 10, 8, 6)$ lattices,
each with three lattice spacings $a \sim (0.064, 0.068, 0.075) $~fm.
The chiral symmetry in the HMC simulation is preserved with $N_s=16$ in the fifth dimension,
and the optimal weights $\{\omega_s, s=1,\cdots, 16\}$ are computed with
$\lambda_{min} = 0.05 $ and $ \lambda_{max} = 6.2 $, with the error
of the sign function of $H_w$ less than $1.2 \times 10^{-5} $, 
for eigenvalues of $ H_w $ satisfying $ \lambda_{min} \le | \lambda(H_w) | \le \lambda_{max} $.
The residual masses of $(u/d, s, c)$ quarks are less than 
(0.09, 0.08, 0.04)~MeV/$c^2$ respectively (see Table \ref{tab:15_ensembles_mres}).
The bare quark masses and lattice spacings are determined on the $64^4$ lattices 
(see Table \ref{tab:a_qmass}). 
For each lattice spacing, the bare quark masses of $u/d$, $s$ and $c$ are tuned
such that the lowest-lying masses of the meson operators
$ \{ \bar{u} \gamma_5 d, \bar{s} \gamma_i s, \bar{c} \gamma_i c \} $
are in good agreement with the physical masses of 
$\{ \pi^{\pm}(140), \phi(1020), J/\psi(3097) \}$ respectively. 
%For hadronic observables, 
%this set of gauge ensembles should give fairly accurate theoretical predictions, 
%in view of the residual masses of $u/d$, $s$ and $c$ quarks 
%less than 0.09, 0.08, and 0.04 (in units of MeV/$c^2$) respectively.

In this paper, we determine the topological susceptibility for $T > T_c$. 
The topological charge of each gauge configuration is measured   
by the clover charge in Wilson flow at the flow time $ t = 0.8192~{\rm fm}^2 $, 
where the topological susceptibility of any gauge ensemble attains its plateau.
Using the topological susceptibility $\chi_t(a,T)$ of 15 gauge ensembles 
with 3 different lattice spacings and different temperatures in the range 
$T \sim 155-516$~MeV (see Table \ref{tab:15_ensembles}),  
we fit the data points to the ansatz (\ref{eq:ansatz}), 
and obtain the fitted parameters and $\chi_t(T)$ in the continuum limit 
(see Fig. \ref{fig:chit14_15pts_a0}).  
In the limit $ T \gg T_c $, it gives $ \chi_t(T) = 12.8(1) (T_c/T)^{8.1(2)} $ 
(in units of ${\rm fm}^{-4}$),
which agrees with the temperature dependence of $\chi_t(T) $ in
the dilute instanton gas approximation (DIGA) \cite{Gross:1980br}, 
$ \chi_t(T) \sim T^{-8.3} $ for $N_f=4$.
This implies that our data points of $ \chi_t(a, T) $ for $ T > 350 $~MeV are valid,
up to an overall constant factor.

To investigate the volume dependence of topological susceptibility, we 
generate another set of ensembles of lattice sizes $ 32^3 \times (16, 12, 10, 8, 6) $,  
with lattice spacing $ a \sim 0.0641~\fm $ and volume $ \sim (2.053~\fm)^3 $, 
and obtain the $\chi_t(a,T) $ for $ T \sim 192-512$~MeV (see Table \ref{tab:5_ensembles}).  
Comparing the topological susceptibilites of this relatively smaller volume
with their counterparts of a larger volume $ \sim (4.074~\fm)^3 $ 
on the $ 64^3 \times (16, 12, 10, 8, 6) $ lattices with lattice spacing $ a \sim 0.0636~\fm $  
in Table \ref{tab:15_ensembles}, and also with those in the infinite volume limit 
by extrapolation (see Fig. \ref{fig:chit14_l64_l32}),  
we conclude that the values of $ \chi_t^{1/4} $ in Table \ref{tab:15_ensembles} 
do not suffer from significant finite-volume systematics. 

Since our present simulation is not at the exact chiral symmetry limit, we investigate the 
feasiblity of using the reweighting method to obtain $ \chi_t $ in the exact chiral 
symmetry limit. In Section \ref{section:reweighting}, we give a detailed discussion of 
the reweighting method for DWF. 
We find that the prerequisite for the reweighting method to work efficiently for DWF
is that the index of the effective 4D Dirac operator $ D_{N_s}(0) $ 
[see Eq. (\ref{eq:Dmq_Ns})] of the DWF is equal to the index of 
the overlap Dirac operator $ D_\ov(0) $ for each configuration in the ensemble. 
Moreover, the approximate sign function $ S_{N_s}(H) $ is required 
to be sufficiently precise, especially for the low-lying eigenvalues of $ |H| $   
(where $ H= H_w = \gamma_5 D_w $ for optimal DWF, 
 and $ H=\gamma_5 D_w(2 + D_w)^{-1} $ for Shamir/M\"obius DWF). 
%which is equivalent to ensure that the chiral symmetry of $ D_{N_s}(0) $ is sufficiently precise. 
Then the weight factor (\ref{eq:W}) of each configuration is of the order one, i.e., $ W_i \sim 1 $. 
To fullfil above requirements, for optimal DWF simulation, it has to use a sufficiently small 
$ \lambda_{min} $ and also a sufficiently large $ N_s $, 
while for the Shamir/M\"obius DWF simulation, 
it needs to use some $ N_s $ much larger than that of optimal DWF, since it does not have 
any parameter like $ \lambda_{min} $ to enhance the chiral symmetry 
for the low-lying eigenvalues of $ | H | $. 

The above approach of obtaining $\chi_t(T) $ in the exact chiral symmetry limit
is to reweight $ \chi_t(a,T) $ to $ {\chi_t}^\ov(a,T) $ at finite lattice spacing $a$ 
and temperature $T$, then use a set of data points of $ {\chi_t}^\ov(a,T) $ 
at many different $a$ and $T$ to extract $ {\chi_t}^\ov(T) $ in the continuum limit for any $T$. 
Nevertheless, this rigorous approach seems to be prohibitively expensive. 
Besides the very expensive DWF simulation with $ \lambda_{min} \lesssim 0.00001 $ 
and $ N_s \gtrsim 32 $, there are even more costly projections of low modes of 
$ V_{N_s} $ and $ V_\ov $ for computing the weight factor of each configuration in the ensembles.
To pursuit this approach is out of question unless the exaflop computers are available. 
Nevertheless, from the viewpoint of universality, 
even $ {\chi_t}^{\DWF}(a,T) $ (measured by the index of $ D_{N_s}(0) $)
at finite lattice spacing is different from $ {\chi_t}^\ov(a,T) $, theoretically,   
in the continuum limit, both $ {\chi_t}^{\DWF}(T) $ 
and $ {\chi_t}^\ov(T) $ should go to the same universal value, 
since both $ D_{N_s}(0) $ and $D_\ov(0)$ go to the massless Dirac operator 
in the continuum limit, provided that the chiral symmetry 
of DWF is sufficiently precise such that 
$ {\rm index} \{ D_{N_s}(0) \} =  {\rm index} \{ D_\ov(0) \} $.  
Then the reweighting at finite lattice spacing seems to be unnecessary. 
Moreover, for an ensemble generated by DWF simulation with effective 4D Dirac operator 
$ D_{N_s}(0) $ sufficiently close to $ D_\ov(0) $, the topological charge of each 
configuration can be measured by $ Q_\clover $ in the Wilson flow, since it is expected 
that $ \chi_t(\clover) = \chi_t ({\rm overlap}) $ in the continuum limit 
and in the infinite volume limit.      
Further studies are required to examine whether any of above scenarios 
could be realized in practice. 

%\eject 

\vfill
\section*{Acknowledgement}

We are grateful to Academia Sinica Grid Computing Center (ASGC)
and National Center for High Performance Computing (NCHC) for the computer time and facilities.
This work is supported by the Ministry of Science and Technology
(Grant Nos.~108-2112-M-003-005, 109-2112-M-003-006, 110-2112-M-003-009).

\eject

\appendix

\section{Renormalized chiral condensate}

As discussed in Section \ref{sec:introduction}, the chiral condensate $ \Sigma $ 
is the order parameter of spontaneously chiral symmetry breaking in QCD.
At low temperatures ($ T < T_c $), the chiral symmetry of QCD is spontaneously broken 
and the chiral perturbation theory (ChPT) gives the relation between the chiral condensate 
and the topological susceptibility. However, at high temperatures ($ T > T_c $), 
the chiral symmetry of $u$ and $d$ quarks is effectively restored and the chiral condensate 
vanishes. Thus the ChPT becomes inapplicable for $ T > T_c $, 
and it cannot give an expression of $ \chi_t(T) $.   
Theoretically, the topological susceptibility can be nonzero for $T > T_c$ due to 
the quantum flucatations of the physical $(u,d,s,c,b)$ quarks with nonzero masses. 
Even though the chiral condensate $\Sigma(T) $ seems to be irrelevant to  
$ \chi_t(T) $ for $ T > T_c $,  
it is important to find out how the chiral condensate $ \Sigma(T) $ changes 
with respect to $T$, and from which to determine the pseudocritical temperature $ T_c $, 
and to understand the nature of the chiral symmetry restoration. 
To this end, lattice QCD provides a viable framework for nonperturbative determination 
of $ \Sigma(T) $ from the first principles.  
   
In lattice QCD, the quark condensate $ \Sigma(m_q,T) $ suffers 
from the quadratic divergences 
%($\sim m_q/a^3 $) 
in the continuum limit. 
To remove the quadratic divergences of the $u/d$ quark condensate, 
one considers the subtracted quark condensate 
\bea
\label{eq:Delus}
\Delta_{us}(T) = \Sigma(m_u,T) - \frac{m_u}{m_s} \Sigma(m_s,T),   
\eea 
where $ m_u $ and $ m_s $ are the bare masses of the $u$ and $s$ quarks.
Moreover, the multiplicative renormalization factor in (\ref{eq:Delus})   
can be eliminated by normalization with its corresponding value at $T=0$, i.e.,    
\bea
\label{eq:DelusR}
\Delta_{us}^R(T) = \frac{ \Sigma(m_u,T) - \frac{m_u}{m_s} \Sigma(m_s,T)} 
                        { \Sigma(m_u,0) - \frac{m_u}{m_s} \Sigma(m_s,0)},  
\eea 
which is called the renormalized chiral condensate \cite{Cheng:2007jq}.

\begin{table}[h!]
\begin{center}
\caption{The renormalized chiral condensate $ \Delta_{us}^R(T)$ of 14 gauge ensembles.}   
\setlength{\tabcolsep}{4pt}
\vspace{2mm}
\begin{tabular}{|ccccccc|}
\hline
    $\beta$ 
  & $a$[fm]
  & $ N_x $
  & $ N_t $
  & $T$[MeV] 
  & $N_{\rm confs}$ 
  & $\Delta_{us}^R(T)$
\\
\hline
\hline
6.15 & 0.0748 & 64 & 20 & 131 & 152 & 0.594(63) \\
6.18 & 0.0685 & 64 & 20 & 144 & 182 & 0.299(34) \\
6.20 & 0.0636 & 64 & 20 & 155 & 218 & 0.133(16) \\ 
6.20 & 0.0636 & 64 & 16 & 193 & 395 & 0.040(8) \\ 
6.18 & 0.0685 & 64 & 12 & 240 & 194 & 0.015(1) \\ 
6.15 & 0.0748 & 64 & 10 & 263 & 274 & 0.010(1) \\ 
6.18 & 0.0685 & 64 & 10 & 288 & 263 & $ 6.58(33) \times 10^{-3} $ \\ 
6.20 & 0.0636 & 64 & 10 & 310 & 243 & $ 2.29(62) \times 10^{-3} $  \\ 
6.15 & 0.0748 & 64 & 8  & 329 & 323 & $ 4.43(16) \times 10^{-3} $  \\ 
6.18 & 0.0685 & 64 & 8  & 360 & 365 & $ 3.39(14) \times 10^{-3} $ \\ 
6.20 & 0.0636 & 64 & 8  & 387 & 317 & $ 2.24(72) \times 10^{-3} $ \\ 
6.15 & 0.0748 & 64 & 6  & 438 & 303 & $ 2.43(1.26) \times 10^{-3} $ \\ 
6.18 & 0.0685 & 64 & 6  & 479 & 382 & $ 0.76(2.42) \times 10^{-3} $  \\ 
6.20 & 0.0636 & 64 & 6  & 516 & 732 & $ 8.27(56) \times 10^{-4} $ \\ 
\hline
\end{tabular}
\label{tab:14_ensembles}
\end{center}
\end{table}

\begin{figure}[!ht]
\begin{center}
\includegraphics[width=11cm,clip=true]{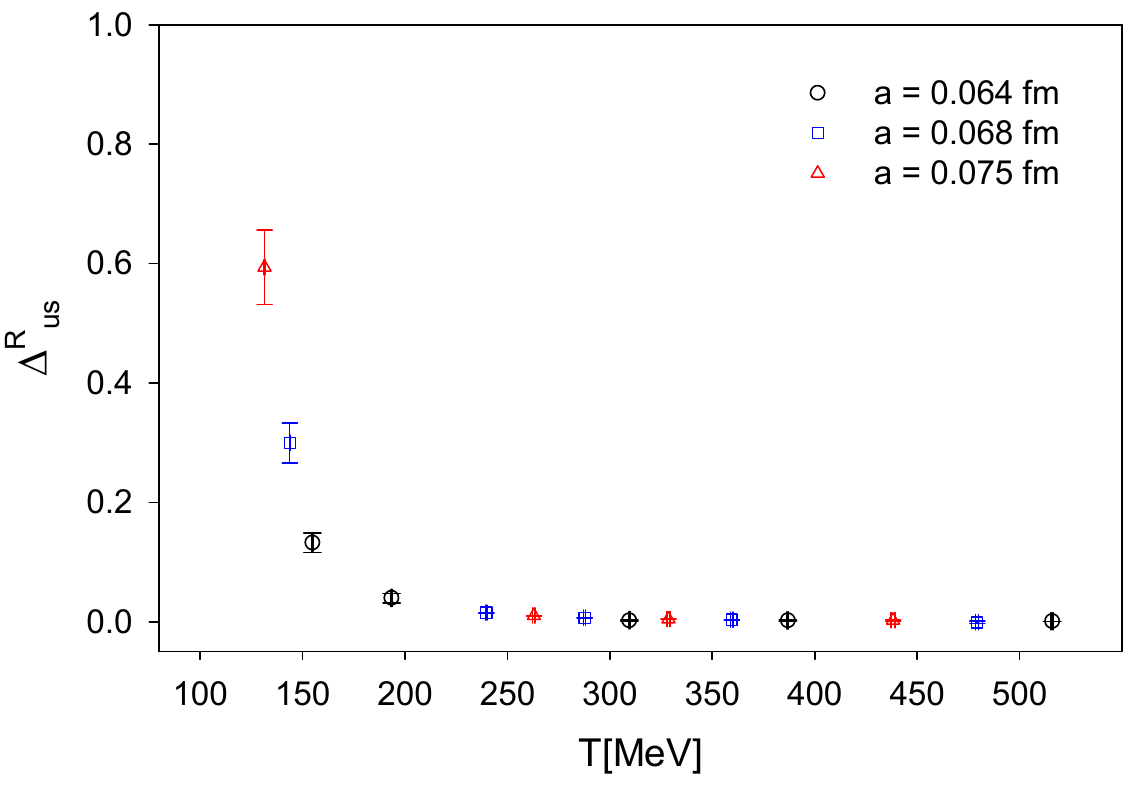}
\caption{The renormalized chiral condensate $\Delta^R_{us}$ versus the temperature $T$.} 
\label{fig:DelusR_l64}
\end{center}
\end{figure}

In the following, we present our first results of $\Delta_{us}^R(T)$ for $T \simeq 131 - 516 $~MeV,  
in lattice QCD with $N_f=2+1+1$ domain-wall quarks at the physical point,  
for the 14 ensembles listed in Table \ref{tab:14_ensembles}. 
There are 12 ensembles with $ T > 150 $~MeV 
(a subset of the ensembles in Table \ref{tab:15_ensembles}),  
and two ensembles with $ T < 150 $~MeV. 
The quark condensate $ \Sigma(m_q, T) = \Tr (D_c + m_q)^{-1}/V$ 
is estimated by the noise method, using 24-240 $Z_2$ noise vectors to evaluate 
the all-to-all quark propagators for each configuration. 
The number of configurations of each ensemble for the evaluation of $ \Delta_{us}^R(T)$ 
is given in the column $N_{\rm confs}$.  The data of $\Delta_{us}^R(T)$ in the last column 
is plotted versus $T$ in Fig. \ref{fig:DelusR_l64}. 
Here the normalization factors (i.e., the denominator in (\ref{eq:DelusR})) for 3 different 
lattice spacings are evaluated on the $64^4$ lattices with (100, 67, 84) configurations for 
$\beta = (6.20, 6.18, 6.15)$ respectively.  

Comparing the data of $\Delta_{us}^R$ in Fig. \ref{fig:DelusR_l64} 
with its counterpart in $N_f = 2+1+1$ QCD with maximally twisted mass Wilson fermions at $ m_\pi = 210$~MeV 
(see Fig. 3 in Ref. \cite{Burger:2018fvb}), we find that the trends of  
the falling of $ \Delta_{us}^R(T) $ (as $T$ is increased from low to high temperatutes)
are consistent with each other.   
However, the fastest falling of $\Delta_{us}^R(T) $ in  
Fig. \ref{fig:DelusR_l64} is in the range of $T \sim 145-150$~MeV,    
while it is $\sim 161-165$~MeV in Ref. \cite{Burger:2018fvb}. 
Moreover, $\Delta_{us}^R(T)$ in Fig. \ref{fig:DelusR_l64} falls steeper than 
its counterpart in Ref. \cite{Burger:2018fvb}.  
The discrepancies can be attributed to the different parameters in these two studies, 
namely, the unphysical $m_\pi = 210$~MeV and a smaller volume $\sim (3.1~{\rm fm})^3$ 
in Ref. \cite{Burger:2018fvb}, versus the physical $m_\pi = 140$~MeV and a larger volume 
$ \sim (4.1~\text{fm})^3 $ in this study. 
In general, as $ m_\pi $ gets smaller, 
the pseudocritical temperature $T_c$ becomes lower, the range of $T$ for   
the fastest falling of $\Delta_{us}^R(T) $ becomes narrower, and the falling becomes steeper. 
Furthermore, these effects are enhanced as the volume gets larger.
 
Comparing the data of $\Delta_{us}^R$ in Fig. \ref{fig:DelusR_l64} with its counterpart
in $N_f = 2+1$ QCD with staggered fermions at the physical point 
and in the continuum limit \cite{Borsanyi:2010bp,Bazavov:2011nk}, 
we find that the fastest falling of $\Delta_{us}^R(T) $ in $N_f = 2+1 $ QCD  
is in the range of $T \sim 155-165$~MeV, while it is $\sim 145-150$~MeV in this study.       
This seems to suggest that the pseudocritical temperature in $N_f=2+1$ QCD 
is $T_c \sim 155-160$~MeV, higher than that ($T_c \sim 145-150$~MeV) in $N_f=2+1+1$ QCD. 
Note that the ratio $ (m_{ud}/m_s)^{\rm phys} $ (the physical point) 
was set to $ \sim 0.0355$ in Ref. \cite{Borsanyi:2010bp} and was extrapolated to 0.037   
in Ref. \cite{Bazavov:2011nk}, which are larger than the $ (m_{ud}/m_s)^{\rm phys} $ values      
$\sim 0.0310-0.03125 $ (see Table \ref{tab:a_qmass}) in this study.
The difference of the ratios of $ (m_{ud}/m_s)^{\rm phys} $ between 
the $N_f = 2 + 1 $ QCD in Refs. \cite{Borsanyi:2010bp,Bazavov:2011nk} and the $N_f = 2+1+1$ QCD in this study
may shed light on the question why $T_c$ in $N_f=2+1+1$ QCD is lower than that in $N_f=2+1$ QCD.
We will return to this question after more data points around $T_c$ will be available 
and a more precise determination of $T_c$ will become feasible.

To determine the pseudocritical temperature $T_c$  
requires many data points of $\Delta_{us}^R(T) $ in the vincinity 
of $ T_c $, which is very challenging and beyond the scope of this paper.    
Note that the HMC simulations of the ensembles at $ T < 150 $~MeV for $N_f=2+1+1$ lattice QCD 
with domain-wall quarks at the physical point are almost prohibitively expensive for us.
Moreover, to get a good estimate of the all-to-all quark propagators by the noise method, 
it is necessary to use a sufficiently large number of noise vectors,  
which turns out to be very computationally intensive 
and requires a large amount of disk space. 
Here we only measure the renormalized chiral condensate of 14 ensembles 
for $ T \sim 130-516 $~MeV, and to see how it decreases to zero as $T$ is increased.
One thing for sure is that all data points of $\chi_t^{1/4}(a,T)$ in Table \ref{tab:15_ensembles} 
are in the chirally symmetric phase.

\end{document}